\documentclass{pasa}%
\usepackage{graphicx,float,placeins}
\usepackage{lipsum,rotating}
\usepackage{multirow}
\usepackage{multicol,caption}
\usepackage{pdflscape}
\usepackage{amsmath}	
\usepackage{amssymb}
\usepackage{amsmath,upgreek}
\usepackage{booktabs}

\title[Isolated Field Elliptical Galaxies from the CFHTLS-W1]{Identification and Properties of Isolated Field Elliptical Galaxies from CFHTLS-W1}
\author[Ulgen et al.]{E. Kaan Ulgen$^1$, Sinan Alis$^{2,3}$, Christophe Benoist$^4$, F. Korhan Yelkenci$^2$, Oguzhan Cakir$^2$, Suleyman Fisek$^2$, and Yuksel Karatas$^2$
\affil{$^1$Astronomy and Space Sciences Program, Institute of Graduate Studies in Sciences, Istanbul University, 34116, \\
Istanbul, Turkey}
\affil{$^2$Department of Astronomy and Space Sciences, Faculty of Science, Istanbul University, 34116 Istanbul, Turkey}
\affil{$^3$Istanbul University Observatory Research and Application Centre, 34116 Istanbul, Turkey}
\affil{$^4$Laboratoire Lagrange, UMR 7293, Universit\'e C\^{o}te d'Azur, CNRS, Observatoire de la C\^{o}te d'Azur, 06304 Nice, France}
}%

\jid{PASA}
\doi{10.1017/pas.\the\year.xxx}
\jyear{\the\year}

\usepackage{aas_macros}
\usepackage{hyperref} 
\hypersetup{colorlinks,citecolor=blue,linkcolor=blue,urlcolor=blue}

\begin{document}

\begin{frontmatter}
\maketitle

\begin{abstract}
We present a catalogue of isolated field elliptical (IfE) galaxies drawn from the W1 field of the Canada-France-Hawaii Telescope Legacy Survey (CFHTLS). 228 IfEs were identified from a flux-limited (r<21.8) galaxy catalogue which corresponds to a density of 3 IfE/sq.deg. For comparison we consider a sample of elliptical galaxies living in dense environments, based on identification of the brightest cluster galaxies (BGCs) in the same survey. Using the same dataset for the comparison sample ensures a uniform selection, including in the redshift range as IfEs (i.e. 0.1 < z < 0.9). A comparison of elliptical galaxies in different environments reveals that IfEs and BCGs have similar behaviours in their colours, star formation activities, and scaling relations of mass-size and size-luminosity. IfEs and BCGs have similar slopes in the scaling relations with respect to cluster ellipticals within the $-24 \leq M_{r} \leq -22$ magnitude and $10.2< \textrm{log}( \textrm M_{*}/ \textrm M_\odot)\leq12.0$ mass ranges. Three IfEs identified in this study can be associated with fossil groups found in the same survey area which gives clues for future studies.
\end{abstract}

\end{frontmatter}

\section{INTRODUCTION}
\label{sec:intro}
Galaxies can be found in different environments, such as clusters, groups, fields, and voids. These regions show relatively high (clusters and groups) and low (fields and voids) galaxy densities. These different environments are populated by galaxies of different types. Specifically, elliptical galaxies are mostly seen in high-density environments. This is characterized by the morphology-density relation shown by \cite{Dressler1980}. However, elliptical galaxies may also be found in low density environments \citep{Smith2004,Reda2004,Hernandez2008,Lacerna2016}. These elliptical galaxies in the field are called isolated field elliptical (IfE) galaxies and their formation scenario is still debated.

While IfEs represent the least dense environment for an elliptical galaxy, the brightest cluster galaxies (BCGs) are giant ellipticals and are found in the densest environments. Thus, BCGs constitute a very special class of galaxies \citep{VonDerLinden2007}. They reside in the core of galaxy clusters and most of the time they are located in the very centre. Their uniqueness can be used as a tool to identify galaxy clusters \citep{Koester2007a,Koester2007b,Hao2010}. The core of a cluster is the densest environment and was shown to play an important role in galaxy evolution \citep{Goto2003,Balogh2004,Bell2006,Ma2011,Conselice14}.

Comparing the properties of galaxies in different environments is crucial to understand their formation and evolutionary paths. Several authors studied the mass-size relation of galaxies in different environments such as cluster and field \citep{Huertas-Company13A,Poggianti13,Delaye14,Kelkar15} and suggested that the size evolution of galaxies is mass-dependent. However, when elliptical galaxies in different environments were compared, \cite{Kelkar15} and \cite{Huertas-Company13B} found no difference in mass-size relations.

Size evolution or size-luminosity relation of BCGs are investigated in many studies; both based on observations \citep{Fasano2010,Ascaso2011,Lidman2012,Lidman2013,Radovich20} and based on simulations \citep{DeLucia2007}.

Various formation scenarios for isolated field ellipticals have been proposed. Since processes seen in high-density environments (e.g. galactic cannibalism, ram-pressure stripping, tidal stripping) cannot be responsible for the formation of IfEs, the most plausible mechanisms are the major merging of galaxies and the collapse of fossil groups. Following \cite{Toomre1972} elliptical galaxies are believed to be formed by the merging of two spiral galaxies. This idea is consistent with the fact that IfEs are found in low-density environments. On the other hand, numerical simulations have shown that compact galaxy groups can be merged into a central elliptical galaxy \citep{Barnes1989}. These galaxies may keep their X-ray emitting gas halo even after the merging event. Such galaxies have been identified by X-ray studies \citep{Ponman1994,Mulchaey1999} and their X-ray luminosities are comparable to those of compact groups. Features such as isolation and the presence of large extended X-ray haloes support the idea that IfEs are related to fossil groups. \cite{Niemi2010} questioned this link between those classes as their formation timescales are found to be different in the Millennium Simulation \citep{Springel2005} but emphasized the similarities between IfEs and fossil groups.

Different definitions have been used in the literature for selecting IfEs \citep{Aars2001,Reda2004,Smith2004,Lacerna2016}. The difference mainly occurs both in the selection of galaxies and in the isolation criteria. In some studies IfE were selected from galaxy catalogues by using various morphological parameters, in others, authors draw their sample from early-type galaxies including lenticulars. Even though the criteria used for detecting IfEs are diverse, the main idea remains the same, central galaxies with less crowded surroundings or with faint companion galaxies.

\cite{Aars2001} identified nine IfE from the catalogue of \cite{Karachentseva1973} and compared their environments with those of loose groups. They showed that half of their IfEs have similarities with loose groups in terms of environmental densities. 

\cite{Colbert2001} compared 23 isolated early-type galaxies with those in poor groups. They showed that the majority of IfEs have dust and tidal features, as well as shells. With the assumption of those features might be produced by merging events, they concluded that IfEs might have experienced these events relatively recently. They found dust features with a similar fraction in both isolated and cluster early-types. However, shells and tidal features are more dominant in IfEs than their cluster counterparts.

\cite{Smith2004} identified 32 IfEs from the LEDA catalogue and found a significant dwarf galaxy population around IfEs.

\cite{Reda2004} examined 36 IfEs and compared them with cluster ellipticals. Their findings indicate a common formation epoch for elliptical galaxies in low and high densities (i.e. isolated and cluster member). Both samples show similar colour distributions, surface brightness properties, and red-sequence slopes in colour-magnitude diagrams. As they also noted some structural features like dust lanes, envelopes, and boxy discs, the formation scenario of a major merger or group collapse seems more plausible. \cite{Lacerna2016} investigated the colours and star-formation properties of 89 IfEs identified by \cite{Hernandez2010} and showed that both isolated ellipticals and Coma cluster ellipticals present similarities. Since elliptical galaxies in both categories are "red" and "dead", they concluded that the colour and hence the morphology of elliptical galaxies are independent of the environment. \cite{Niemi2010} studied the 293 IfEs identified from the Millennium Simulation. They compared IfEs with elliptical galaxies that are not in isolation. In contrast to observations, simulated IfEs are on average bluer than non-isolated ellipticals. In addition, almost half of their IfE sample shows the presence of major merger events whereas almost all IfEs experience at least one merging event (e.g. minor merger). Only one-third of their non-isolated elliptical galaxies show a major merging activity in their merger histories.

In this study, we identified IfEs in a flux-limited galaxy catalogue obtained from the W1 field of the Canada-France-Hawaii Telescope Legacy Survey (CFHTLS) \citep{Hudelot2012} and compared their basic properties with cluster ellipticals and BCGs as representative samples of high-density environments. For the first time, a sample of IfEs is compared with a similar sample of BCGs in the same redshift regime.
 
The structure of the paper is as follows. The survey, data, and sample selection are described in Section 2; brightness, colour, star formation activity, and environmental density of IfEs are compared with BCGs and cluster ellipticals in Section 3; scaling relations for size-luminosity and mass-size are given in Section 4; possible fossil group connection is given in Section 5, a discussion on the results is given in Section 6, and the summary of our study is given in Section 7. Throughout this paper we use H$_{0}$ = 70 km s$^{-1}$ Mpc$^{-1}$, $\Omega_{ \textrm m}$ = 0.3 and $\Omega_{\Lambda}$ = 0.7. 

\section{Data and Sample Selection}

\subsection{CFHTLS Data}

The data used in this work were obtained from the seventh and the final data release of the CFHTLS (T0007, 2012) which is covering a total of 155 deg$^2$.  The CFHTLS was carried out using MegaCam, a mosaic imager, mounted at the 3.6 m CFH Telescope and completed in 2009, is an optical five bands ($u*, g', r', i'/y', z'$) survey consisting of four wide and deep fields. MegaCam has 36 2048x4612 pixels CCDs with a pixel scale of 0.186"/pixel. Thus, each MegaCam pointing yields a field of view of $\sim$ 1 deg$^2$. In this work, we used CFHTLS W1 (72 deg$^2$, 8$^\circ$ x 9$^\circ$), the widest field of the survey, to identify isolated field ellipticals. Coordinates and limiting magnitudes of the CFHTLS-W1 can be seen from Table \ref{tabsurvey}.

\begin{table}[ht]
\caption{Equatorial and galactic coordinates of the CFHTLS-W1, limiting magnitudes for extended objects in each photometric band and the total effective survey area are given. Other major extragalactic surveys in the same field are listed in the last column.} 
\centering
\begin{tabular}{|c|c|}
\hline
$\alpha$ ($^{\circ}$) &34.500\\
$\delta$ ($^{\circ}$) &-7.000 \\ 
$l$  ($^{\circ}$) &172.468 \\
$b$  ($^{\circ}$) &-61.242 \\ 
$u$ (mag) &24.5 \\
$g$ (mag) &24.7 \\
$r$ (mag) &24.0\\    
$i$ (mag) &23.7\\
$z$ (mag) &22.9\\
area (sq.deg) &72\\
Related Surveys &XXL, VIPERS, GAMA\\
\hline
\end{tabular}\label{tabsurvey}
\end{table}

The CFHTLS raw data were reduced and were made public by the former laboratory TERAPIX of the IAP\footnote{http://www.iap.fr}. Image stacking and astrometry were performed using SWARP \citep{Bertin2006} and photometry was performed with SExtractor \citep{Bertin1996}. Data released by TERAPIX includes images, object catalogues, mask files, and photometric redshifts. For the present work, we revisited mask files provided by TERAPIX. Although we updated masks with a more conservative approach, due to the overmasked regions in the original data release, we have more galaxies in our catalogue. The original object catalogue provided by TERAPIX includes 2,623,690 ($ \textrm r \leq24$) galaxies whereas our catalogue includes 2,812,065 ($ \textrm r \leq24$) galaxies. These galaxy counts obtained with all five-band magnitudes ($u*, g', r', i'/y', z'$) are measured by omitting galaxies with missing magnitude in any band. Contents of the object catalogues can be found on relevant pages at CFHT webpage\footnote{https://www.cfht.hawaii.edu/Science/CFHTLS/T0007/}.

Photometric redshifts of the objects were computed using \texttt{LePhare} at TERAPIX and Laboratoire d'Astrophysique Marseille \citep{Ilbert2006,Coupon2009}. Several spectroscopic surveys, such as VVDS and COSMOS, have been used for the calibration of those photometric redshifts. Typical photometric redshift error for elliptical galaxies in the CFHTLS is $ \upsigma_{ \textrm z}=0.03\times(1+ \textrm z)$ \citep{Coupon2009}. The bias shown by \cite{Coupon2009} is mostly at the lowest ($ \textrm z<0.1$) and highest ($ \textrm z>0.9$) redshift regimes. But for our sample (i.e. elliptical galaxies) we did not apply any correction to the redshift bias as it is negligible for elliptical galaxies in the range of our study.

\begin{figure}
\begin{center}
	\includegraphics[width=8 cm]{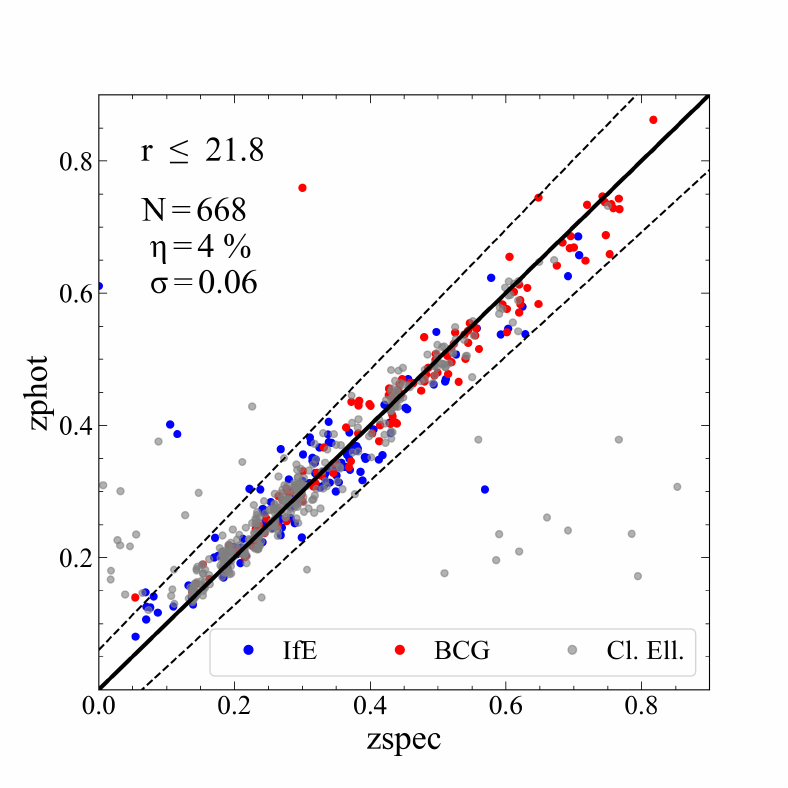}
	\caption{Comparison of photometric redshifts used in this study with corresponding spectroscopic redshifts. Spectroscopic redshifts were compiled from  SDSS, GAMA, and VIPERS galaxy redshift surveys. Blue, red, and gray points represent IfEs, BCGs, and cluster ellipticals, respectively.}
	\label{zpzs}
\end{center}
\end{figure}

\cite{Coupon2009} provides a detailed comparison of the photometric redshifts with spectroscopic redshifts obtained from various sources. Here, we only provide the $zspec-zphot$ comparison for our galaxy sample used for the analysis in Section 3. Fig. \ref{zpzs} shows the comparison for 668 galaxies from our sample where we could compile spectroscopic redshifts from the surveys SDSS, GAMA, and VIPERS. It is seen from the figure that the photometric redshifts of our sample galaxies are in good agreement with spectroscopic ones. The outlier fraction of $\upeta = 4\%$ shown in Fig. \ref{zpzs} is comparable for the same magnitude range of \cite{Coupon2009}.

\texttt{LePhare} is a template fitting photometric redshift algorithm based on galaxy spectral energy distributions \citep{Arnouts1999,Ilbert2006}. There are five spectral types (E, Sbc, Scd, Irr, SB) associated with galaxies once the best match is obtained with \texttt{LePhare} during the SED fitting procedure. These spectral types have been adopted from the observed galaxy spectra given by \cite{CWW80} and \cite{Kinney96} and used similarly with \cite{Ilbert2006}. Four spectra from \cite{CWW80} and two spectra from \cite{Kinney96} were extrapolated into 66 spectral templates and optimised specifically for CFHTLS to represent a wide range of galaxies \citep{Coupon2009}. We make use of these associated spectral types given in the photometric redshift catalogue for the galaxy selection.

Due to the different contents of the object catalogue and the photometric redshift catalogue we created a working catalogue for our study by merging the necessary parameters from those two catalogues.

\subsection{Selection Criteria for IfEs}

We selected elliptical galaxies from our working catalogue based on their spectral type given by \texttt{LePhare} and their magnitude. We applied a magnitude cut at $ \textrm r\leq 21.8$ to obtain a flux-limited sample. 90,872 elliptical galaxies remain in our catalogue after the selection. Each elliptical galaxy is checked for the isolation criteria given by \cite{Zaritsky1993,Zaritsky1997} which was also used by \cite{Smith2004} and \cite{Niemi2010}. The isolation criteria require that the magnitude difference between a neighbour galaxy and the IfE candidate must be greater than 0.7 mags for galaxies within a projected distance of 1 Mpc and greater than 2.2 mags within 500 kpc (for a schematic description of the criteria see Fig. 1 in \cite{Smith2004}). To do that, we created a galaxy catalogue that is 2.2 magnitudes fainter than our elliptical galaxy sample. This reference, fainter, galaxy catalogue contains 2,812,065 ($ \textrm r \leq 24$) galaxies as it was mentioned in the previous section. Selection of the elliptical galaxies was made so that the reference galaxy catalogue does not contain galaxies fainter than the r-band limiting magnitude for the extended objects given in Table \ref{tabsurvey}. The morphology of the neighbouring galaxies is not taken into account in our isolation criteria.

To apply the isolation criteria, mini galaxy catalogues for each elliptical galaxy have been created. These mini catalogues were built from galaxies around the IfE candidate with a projected radius of 1.25 Mpc at the photometric redshift of the galaxy. To account for the large photometric redshift errors, we considered that two galaxies may be at the same redshift if the difference of their photometric redshifts satisfies $\Delta{ \textrm z}\leq2\upsigma_{ \textrm z}$. If this criterion is fulfilled, the IfE candidate is excluded from the list.

\begin{figure*}[ht]
\begin{center}
	\includegraphics[width=17cm]{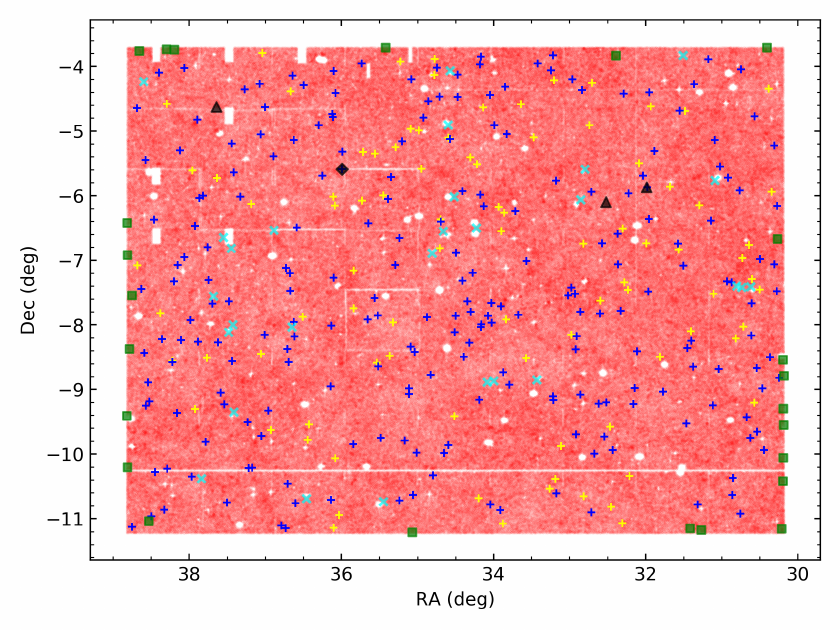}
	\caption{Positions of IfEs shown on the RA-Dec plot. In the figure; red dots are the whole CFHTLS-W1 galaxy catalogue, cyan crosses are 28 IfEs near the masked regions, green squares are 24 IfEs near the field edges, yellow crosses are 89 non-eliptic isolated galaxies, black diamond is a X-ray point source possibly counterpart of an IfE (see Section 5 for details), black triangles are fossil groups identified by \cite{Adami2018} and blue crosses are the final 228 IfEs candidates.}
    \label{positions}
\end{center}
\end{figure*}

Afterwards, for the remaining candidates, with a conservative approach, projected physical distances between the IfE candidates and their neighbouring galaxies are computed at the lowest photometric redshift of each considered pair. Magnitude differences and projected distances are taken into account to check isolation criteria. 369 galaxies satisfying the criteria are identified as isolated. However, some of the IfE candidates are located near the edge of the field (W1) or too close to masked areas around the bright stars, ghosts, or other image defects. To ensure that our IfE identification is not biased by the presence of these regions where the number of galaxies may  artificially be depleted, we checked whether each candidate is closer to the edge by a projected distance of less than 1 Mpc. 24 candidates satisfying this criterion are omitted. Similarly, the fraction of masked areas within a disk of 1 Mpc radius surrounding each IfE candidate was computed. If the masked fraction of the 1 Mpc region around the IfE is higher than 10\% then that IfE is also omitted. 28 candidates in such conditions are excluded from the list. Thus, a total of 317 IfE candidates remained. Positions of those candidates as well as IfEs near masked areas are shown in Fig. \ref{positions}.

Since the selection of IfE candidates is based on measured spectral types, we also removed galaxies classified mistakenly as ellipticals due to the template fitting procedure adopted by \texttt{LePhare}. To do that, we created true-colour image of each isolated galaxy using STIFF\footnotetext{https://www.astromatic.net/software/stiff} \citep{Bertin2012} from $g$, $r$, and $i$ band CFHTLS images. 89 IfE candidates are excluded from the list after a visual inspection of those images. Excluded candidates are in general lenticular galaxies or face-on spirals with a high bulge-to-disk ratio where SED fitting fails.

Thus, 228 isolated field elliptical galaxies remained in our final list. In this way, the purest possible sample is obtained. The density of those 228 IfEs is 3 IfE/sq.deg. One elliptical galaxy in almost every 400 in our magnitude range is an IfE. This makes the IfE fraction in elliptical galaxies 0.25\%. This fraction decreases to $8.1 \times 10^{-3}\%$ when all galaxies (regardless of type) in our working catalogue (i.e. 2,812,065) are considered. Properties of these IfEs are given in Table \ref{tabife}. Colour images of some illustrative IfEs are given in Fig. \ref{Cutout} in the Appendix.

\subsection{Reliability of the IfE Selection}

To test our IfE identification method and to validate the effect of photometric redshift errors, we used a spectroscopic control sample. This control sample is derived from the Galaxy and Mass Assembly (GAMA) Survey. GAMA is a spectroscopic survey to study the evolution of galaxy groups and clusters \citep{Driver2009}. The GAMA Survey was carried out at the Anglo-Australian Telescope using the AAOMega instrument, a high multiplex fiber unit, to cover as many galaxy members as possible of groups and clusters. As a strategy, galaxy groups and clusters are pointed more than once during the GAMA Survey to overcome the difficulties of placing fibres. This strategy leads to high density coverage over the surface area of groups and clusters. GAMA Survey reaches r $<$ 19.8 which corresponds to the redshift regime of z $\leq$ 0.5. A total of 300,000 galaxy spectra have been obtained \citep{Baldry2010,Baldry2014,Robotham2010}.

\begin{figure}
\begin{center}
	\includegraphics[width=\columnwidth]{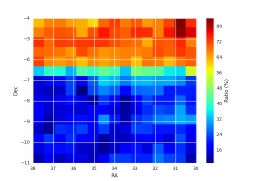}
	\caption{Fraction of CFHTLS-W1 galaxies brighter than \\ $ \textrm r<19.5$ covered by the GAMA-G02 survey.}
	\label{gama}
\end{center}
\end{figure}

Among the survey areas, G02 is the only one overlapping with CFHTLS-W1. We cross-matched galaxies in the G02 field with W1 and plotted them as a function of the match ratio. In Fig. \ref{gama} it can be seen that the GAMA survey is not homogeneous over the entire CFHTLS-W1. The densest area is between $-6^{\circ} < \delta < -4^{\circ}$. There are 33,277 galaxies from G02 spread over the W1 but 17,793 of them lie above roughly $ \delta > -6^{\circ}$. Thus, our control sample was drawn from galaxies belonging to that area in order to test our identification method. CFHTLS-W1 and GAMA-G02 catalogues were cut within the same field $-6^{\circ} < \delta < -4^{\circ}$ and with $ \textrm r\leq19.5$. This leaves 11,657 and 11,282 galaxies in the CFHTLS and GAMA catalogues, respectively. This yields 97\% completeness ($ \textrm r\leq19.5$) for GAMA when compared with CFHTLS. Matching G02 and W1 catalogues led to a galaxy sample that includes galaxy spectral types, magnitudes, and photometric and spectroscopic redshifts. Elliptical galaxies were selected according to their spectral types. Since IfE candidates should be 2.2 magnitudes brighter than the neighbour galaxies, we applied a cut at $ \textrm r\leq17.3$ for ellipticals. Thus, the final control sample consists of 11,282 galaxies ($ \textrm r\leq19.5$) and 130 ellipticals ($ \textrm r\leq17.3$).

When IfE candidates are determined from the control sample using their spectroscopic redshifts, a typical redshift error of $\upsigma_{ \textrm z}$=0.003 is applied. This corresponds to a radial velocity of 1000 km s$^{-1}$ and is considered as a limit for the separation of two galaxies \citep{Colbert2001}. As a result, 38 amongst 130 elliptical galaxies in our control sample were identified as IfE.

Then, we run our selection procedure for the very same galaxies but this time using their photometric redshifts. In this approach, we took into account two aspects: redshift error and implication of this error on the radii we used. Thanks to the spectroscopic sample we can show the impact of photometric redshifts on our IfE identification. For the photometric redshift error we used 1$\upsigma_{ \textrm z \textrm p}$, 2$\upsigma_{ \textrm z \textrm p}$, and 3$\upsigma_{ \textrm z \textrm p}$ where $\upsigma_{ \textrm z \textrm p}=0.03*(1+ \textrm z_{ \textrm c} )$ with $ \textrm z_{ \textrm c}$ being the central galaxy's redshift. To demonstrate the effect on the radius, we use somewhat lower redshifts conservatively. We computed the projected radius with a factor of $\upsigma_{zp}$ smaller than the candidate elliptical galaxy's actual redshift. As a result, to check the isolation criteria, we use the larger, more conservative, radius. We computed these modified radii using redshifts with 1$\upsigma_{ \textrm z \textrm p}$, 2$\upsigma_{ \textrm z \textrm p}$, and 3$\upsigma_{ \textrm z \textrm p}$ smaller than the central galaxy's redshift (i.e. candidate elliptical) as $ \textrm z= \textrm z_{ \textrm c}- \textrm d \textrm z$ where $ \textrm d \textrm z$ is the relevant $\upsigma_{ \textrm z \textrm p}$ . Results of these tests are given in Table \ref{tabcontrol}. $ \textrm r_{ \textrm c \textrm o \textrm r}$ in Table \ref{tabcontrol} denotes these modified (i.e. larger) radii.

\begin{table}[!ht]
\centering
\small{
\caption{Results obtained from the spectroscopic control sample constructed from the GAMA Survey. Six cases listed in the table include different photometric redshift error with and without the usage of modified radii ($ \textrm r_{ \textrm c \textrm o \textrm r}$). In each case, IfEs determined with photometric redshift are matched with the spectroscopically determined 38 IfEs. Number of these matches and their corresponding completeness and purity values are also given.}
\begin{tabular}{|c|c|c|c|c|}
\hline
& zphot & \multicolumn{3}{c|}{zpsec (N:38)}\\
\hline
\multirow{2}{*}{Case} & \multirow{1}{*}{IfE} & \multirow{1}{*}{Match} & \multirow{1}{*}{Completeness} & \multirow{1}{*}{Purity} \\                     
& (N)  &    (N)              &        (\%)                   &    (\%)                      \\
\hline
\multirow{1}{*}{1} & \multirow{2}{*}{21} & \multirow{2}{*}{18}  & \multirow{2}{*}{55} & \multirow{2}{*}{86} \\
$1\sigma_{zp}$ &      &      &  & \\
\hline
\multirow{1}{*}{2} & \multirow{2}{*}{12} & \multirow{2}{*}{10}  & \multirow{2}{*}{32} & \multirow{2}{*}{83} \\
$2\sigma_{zp}$ &      &      &  & \\
\hline
\multirow{1}{*}{3} & \multirow{2}{*}{7} & \multirow{2}{*}{7}  & \multirow{2}{*}{18} & \multirow{2}{*}{100} \\
$3\sigma_{zp}$ &      &      &  & \\
\hline
\multirow{1}{*}{4} & \multirow{2}{*}{11} & \multirow{2}{*}{10}  & \multirow{2}{*}{29} & \multirow{2}{*}{91} \\
$r_{cor}$, $1\sigma_{zp}$ &      &      &  & \\
\hline
\multirow{1}{*}{5} & \multirow{2}{*}{5} & \multirow{2}{*}{4}  & \multirow{2}{*}{13} & \multirow{2}{*}{80} \\
$r_{cor}$, $2\sigma_{zp}$ &      &      &  & \\
\hline
\multirow{1}{*}{6} & \multirow{2}{*}{2} & \multirow{2}{*}{2}  & \multirow{2}{*}{<1} & \multirow{2}{*}{100} \\
$r_{cor}$, $3\sigma_{zp}$ &     &      &  & \\
\hline
\end{tabular}
\label{tabcontrol}
}
\end{table}

Results of the run with photometric redshifts are 21, 12 and 7 IfEs candidates for 1$\upsigma_{ \textrm z \textrm p}$, 2$\upsigma_{ \textrm z \textrm p}$, 3$\upsigma_{ \textrm z \textrm p}$, respectively. When compared with 38 IfEs identified spectroscopically in the same area, 18 (86\%), 10 (78\%), and 7 (100\%) IfEs are found in common. This gives a high purity though the completeness is below 60\%. When this same comparison is done together with the radius correction ($ \textrm r_{ \textrm c \textrm o \textrm r}$) 11, 5, and 2 IfEs are identified, respectively (see Table \ref{tabcontrol}). Balancing sample purity versus completeness, we decided to use 1$\upsigma_{ \textrm z \textrm p}$ error both on photometric redshift and the radius correction (i.e. Case 4 in Table \ref{tabcontrol}). This choice results in a completeness of approximately 30\%, as given in the Table \ref{tabcontrol}.

\subsection{BCGs and Cluster Ellipticals}

To understand environmental effects on elliptical galaxy evolution, we built a BCG sample from the same survey. Thus, we can compare elliptical galaxies in two different and opposite environments; isolated fields and dense regions. Since BCGs may be considered special galaxies in clusters, we also create a sample of elliptical galaxies from the same clusters with our BCG sample.

The galaxy clusters, where BCGs and cluster ellipticals were drawn, were detected in the whole CFHTLS-W1 by the \texttt{WaZP} optical cluster finder. Details of the \texttt{WaZP} cluster finder can be found in \cite{Aguena21,EuclidCol19,Dietrich14}. Basically, the algorithm uses a galaxy catalogue that includes positions (RA and Dec) and photometric redshifts. Overdensities are determined with the help of wavelet filtering and no assumption on the underlying galaxy population is made. The cluster centre is defined by the position of the overdensity peak, which does not always coincide with the BCG. Once an overdensity is detected, it is ranked by the signal-to-noise ratio with respect to the local background of galaxy density. This local galaxy density is computed for the radius and the richness ($\uplambda$) computation of the cluster. Therefore, cluster radius and richness are computed jointly where the radius is determined as 200 times the local galaxy density, and richness is the sum of the membership probabilities within this radius (i.e. $R_{200}$).

There are 3337 cluster detections with a $\textrm S \textrm N \textrm R > 3$ in the whole W1 region. However, to underline the impact of higher density we keep the richest systems in our cluster catalogue by applying a richness cut of $\uplambda \geq 25$. This selection yields 309 galaxy clusters with a median richness of $\uplambda_{ \textrm {med}}\sim35$ whereas the parent cluster catalog in the CFHTLS-W1 has a median richness of $\uplambda_{\textrm{med,all}}\sim10$ (Benoist et al., in prep.). Afterwards, we identify the BCGs of these clusters as briefly explained below.

An elliptical galaxy is identified as a BCG based on the following criteria:

\begin{itemize}  
\item Being 0.5 Mpc around the cluster centre
\item Having a redshift consistent with the cluster with a $\Delta  \textrm z = 0.03*(1+ \textrm z_{ \textrm c \textrm l})$
\item Having a $( \textrm r- \textrm i)$ colour consistent (within $\pm 0.3$) with model elliptical galaxy colours for the corresponding redshift
\end{itemize}

\noindent Then, the BCG is the brightest galaxy in a cylinder defined by the criteria given above.

We select elliptical galaxies from the same clusters as the BCGs. Cluster ellipticals were selected from the core region of clusters with a $ \textrm r < 0.5$ Mpc projected radius using their \texttt{LePhare} defined spectral types (i.e. MOD<23) and photometric redshifts. When selecting cluster ellipticals, the BCGs are excluded in order not to create any bias. These galaxies were also selected based on their membership probability which is computed by \texttt{WaZP} as given in \cite{Castignani16}. A relatively conservative membership probability of $ \textrm P_{ \textrm m \textrm e \textrm m} > 0.7$ is applied for the selection which yields a total of 3958 elliptical galaxies in those 309 rich clusters.

Following the selection of BCGs and cluster ellipticals, we apply the same magnitude cut (i.e. $ \textrm r \leq 21.8$) in order to obtain comparable flux-limited samples with IfEs. Thus, the number of BCGs dropped to 261, and the number of cluster ellipticals dropped to 2087 as the final sample used for the analysis.

In Section 3 we provide a comparison of these BCGs and cluster ellipticals with our IfEs.

\subsubsection{The early-type fraction}

We have not performed a visual inspection for the BCGs and cluster ellipticals as we did for IfEs due to their excess numbers (see Table \ref{tabsample}). In fact, the collective existence of many galaxies in rich clusters makes cluster detection more reliable. Moreover, our cluster elliptical sample is drawn from the cluster core with high membership probabilities. Selecting galaxies from the inner parts of clusters ensures their types as indicated by the morphology-density relation.

Nevertheless, we examined their $M_{u} - M_{r}$ colours to test our selection. Galaxy colours show a bimodal distribution where elliptical galaxies are seen in the red-sequence and are mostly called as \textit{red and dead}. Blue, star-forming galaxies (e.g. spirals or late-types) are found in the blue cloud. There are some galaxies in between these groups called \textit{green-valley} galaxies (see Fig. 2 in \cite{Schawinski2014}). 

\begin{figure}[ht]
\begin{center}
	\includegraphics[width=\columnwidth]{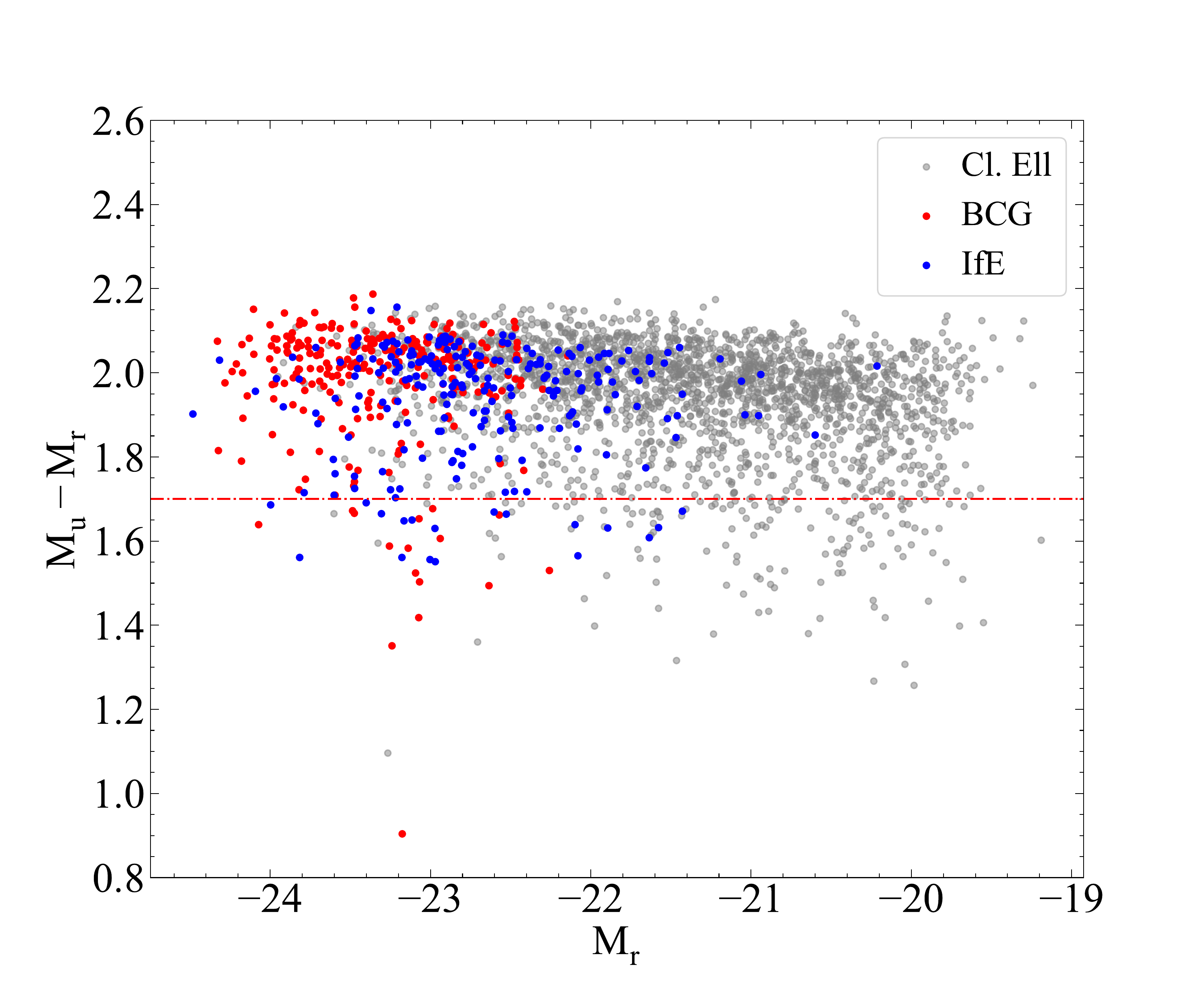}
	\caption{$M_{u}-M_{r}$ colour-magnitude diagram for the galaxy samples. Red-dashed line is the separation of red and blue galaxy populations obtained with GMM at 1.7.}
	\label{cmdur}
\end{center}
\end{figure}

To determine the separation between red and blue galaxies we used member galaxies from clusters with richness $\uplambda > 15$ (i.e. $\sim 1200$ clusters) in the CFHTLS-W1 (Cakir et al., in prep). Thus, we make use of 60,606 cluster members where we implemented a Gaussian-mixture-model (i.e. GMM) which is an unsupervised clustering algorithm. The best separation in the colour space of the two populations was obtained as $(M_{u}-M_{r}) \sim 1.7$ and shown in Fig. \ref{cmdur}. The red-dashed line in the figure denotes this separation. In SDSS, similar exercise was done by \cite{strateva01} and they obtained $(M_{u}-M_{r}) \sim 2.2$ for the separation of blue and red galaxies.

We then apply the obtained separation to $(M_{u}-M_{r})$ colours of our BCGs and cluster ellipticals. 93\% of our BCGs (245 red, 16 blue) and cluster ellipticals (1949 red, 138 blue) are located on the red-sequence. This does not guarantee a correct morphology but still provides a strong indication of their early-type nature.

\subsection{Stellar Masses}

We compute stellar masses for our galaxy sample in order to compare galaxy sizes for a given mass range.

To obtain stellar masses we make use of the Code Investigating GALaxy Emission (\texttt{CIGALE}) \citep{Burgarella05,Noll09,Boquien19} using apparent magnitudes of galaxies in five bands (i.e. ugriz). Galaxy magnitudes were converted to fluxes as \texttt{CIGALE} requires. For the redshift prior we use photometric redshifts provided by TERAPIX.

Stellar masses were computed assuming a single stellar population model of \cite{BC03}, a \cite{Chabrier03} IMF, and the solar metallicity. We adopt \cite{Calzetti00} for dust attenuation law. Then, we run CIGALE for a range of input parameters as $\tau$ = [100-2000] Myr, age=[500-13,000] Myr, and E(B-V)=[0.0-0.2]. Since our sample consists of elliptical galaxies, we use a star formation history with a single burst and a single exponential declining law for SFR as described in \cite{Boquien19}:

\begin{equation}
    SFR(t) \propto \frac{t}{\tau^{2}} \times exp(-t/\tau) \ for \ 0\leq t \leq t_{0}
\end{equation}

\noindent where $\textrm t_0$ is the age of the Universe, $\tau$ is the star formation time-scale (or decay time) and $t$ is the look-back time.

\begin{figure}[ht]
\begin{center}
	\includegraphics[width=\columnwidth]{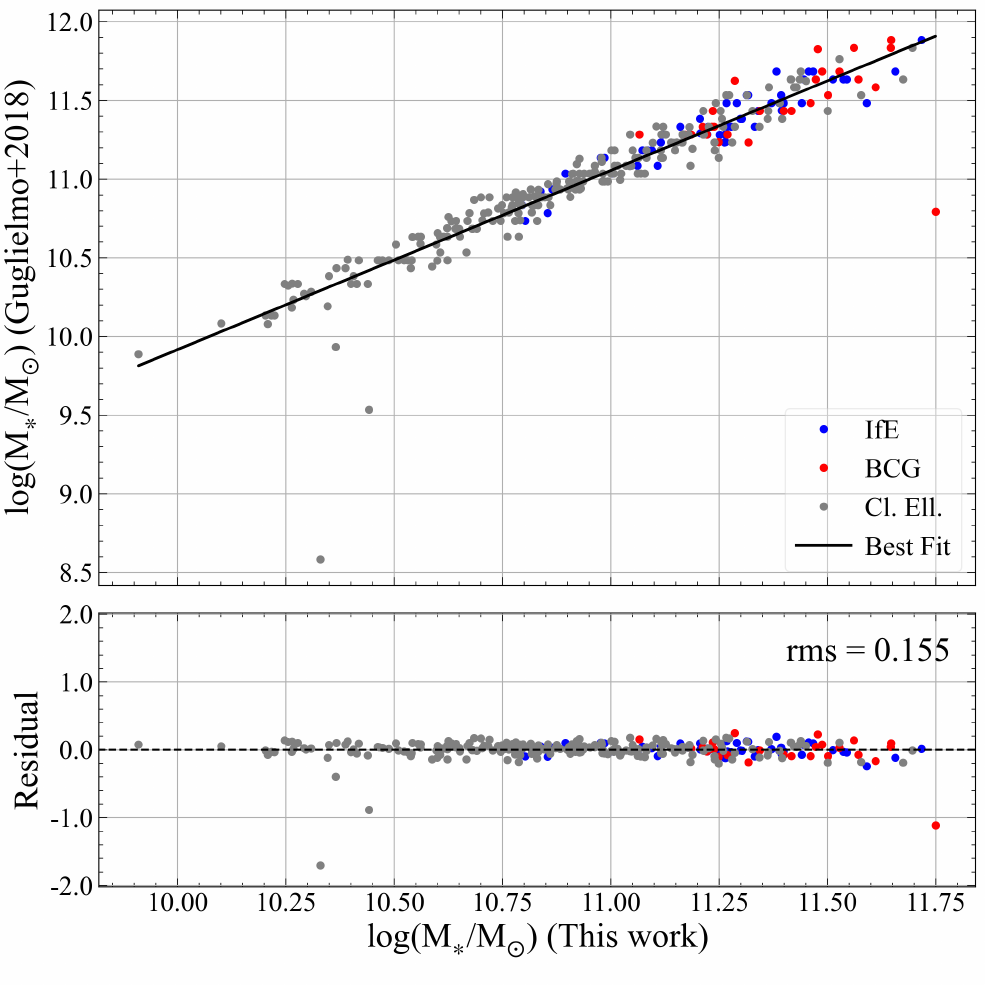}
	\caption{Comparison of stellar masses computed in this study and those in \cite{Guglielmo18}.The rms value of 0.155 of the residuals is comparable with the typical error on the stellar mass we computed in this study. IfEs, BCGs, and cluster ellipticals are denoted by blue, red, and gray points, respectively.}
	\label{masscomp}
\end{center}
\end{figure}

The median value of stellar masses for the whole sample (i.e. 2576 galaxies) given in Table \ref{tabsample} is $\textrm{log} (\textrm M_{med})=10.83 \ M_\odot$, and the median error is 0.10 dex. In order to check our mass estimation, we compared our masses with masses given by \cite{Guglielmo18}. They built a spectrophotometric galaxy catalog for the X-ray detected and spectroscopically confirmed groups and clusters within the northern field of the XXL survey which significantly overlaps with CFHTLS-W1. Their catalog contains 24,336 galaxies with $ \textrm z < 0.6$ and $ \textrm r < 20 $. In total; 2225 member galaxies for 132 groups/clusters, and 22,111 field galaxies. Stellar masses given in their catalogue were computed by using \texttt{LePhare} for the galaxies having at least two observed magnitudes and a spectroscopic redshift. A cross-match of our sample with their catalogue revealed 297 galaxies (52 IFEs, 27 BCGs, and 218 cluster ellipticals) in common. The comparison of the stellar masses is shown in Fig. \ref{masscomp}. Based on the common galaxies, stellar masses from both studies are in good agreement. The dispersion between both estimates is comparable with the typical error of stellar masses computed with \texttt{CIGALE} in this study. The rms value of the one-to-one comparison is 0.155 dex, but it becomes 0.082 dex when we exclude the outliers.

For a detailed comparison of three elliptical galaxy samples in our study, we divided them into four redshift bins. To obtain a reliable comparison and the scaling relations given in Section 4, we also compute our mass completeness limits for each redshift bin by following the approach of \cite{Pozzetti10}, similarly with \cite{Ilbert13,Shi21}.

The recipe we employ for the determination of the mass completeness limits is as follows:

\begin{itemize}
    \item For each redshift bin we identify the 20\% faintest galaxies in the $i-band$ 
    \item For those galaxies we apply the empirical relation given by \cite{Pozzetti10} and compute $ \textrm l \textrm o \textrm g( \textrm M_{ \textrm l \textrm i \textrm m})$
    \begin{equation}
        log(M_{lim}) = log(M_{*}) + 0.4 (i - i_{lim})
    \end{equation}
    \item We then take the 2$\upsigma$ upper envelope of the $ \textrm{log} ( \textrm M_{ \textrm{lim}})$ distribution as the representative limit value
\end{itemize}

Obtained mass completeness limits for the four redshift bins are as follows: $ \textrm{log} ( \textrm M_{ \textrm{lim}})=8.98 \ M_\odot$ for $0.1< \textrm z\leq0.3$, $ \textrm{log} ( \textrm M_{\textrm{lim}})=9.54 \ M_\odot$ for $0.3< \textrm z\leq0.5$, $ \textrm{log} ( \textrm M_{\textrm{lim}})=9.89 \ M_\odot$ for $0.5< \textrm z\leq0.7$, and $ \textrm{log} ( \textrm M_{ \textrm{lim}})=10.39 \ M_\odot$ for $0.7< \textrm z\leq0.9$.

\section{Comparison of IfE\lowercase{s}, BCG\lowercase{s} and Cluster Ellipticals}

In this section, we compare the brightness, colour, environmental density, and star formation activity of IfEs with both a sample of BCGs and a sample of cluster ellipticals described in the Section 2.4.

\begin{table}[h!]
\caption{Description of the galaxy samples used in this study. The table lists number of galaxies ($ \textrm r \leq 21.8$), redshift range and the median redshift for each sample.}
\centering
\begin{tabular}{@{}cccc@{}}
\noalign{\vskip 1mm}
\hline
Sample  & NGAL  &  Redshift Range & Median z \\
\noalign{\vskip 1mm}
\hline
IfE      & 228  & $0.07 \leq \textrm z \leq 0.91$  & 0.32  \\
BCG      & 261  & $0.14 \leq \textrm z \leq 0.86$  & 0.49  \\
Cl. Ell. & 2087 & $0.12 \leq \textrm z \leq 0.86$  & 0.33  \\
\hline
\end{tabular}
\label{tabsample}
\medskip
\end{table}

\begin{figure}
\begin{center}
	\includegraphics[width=\columnwidth]{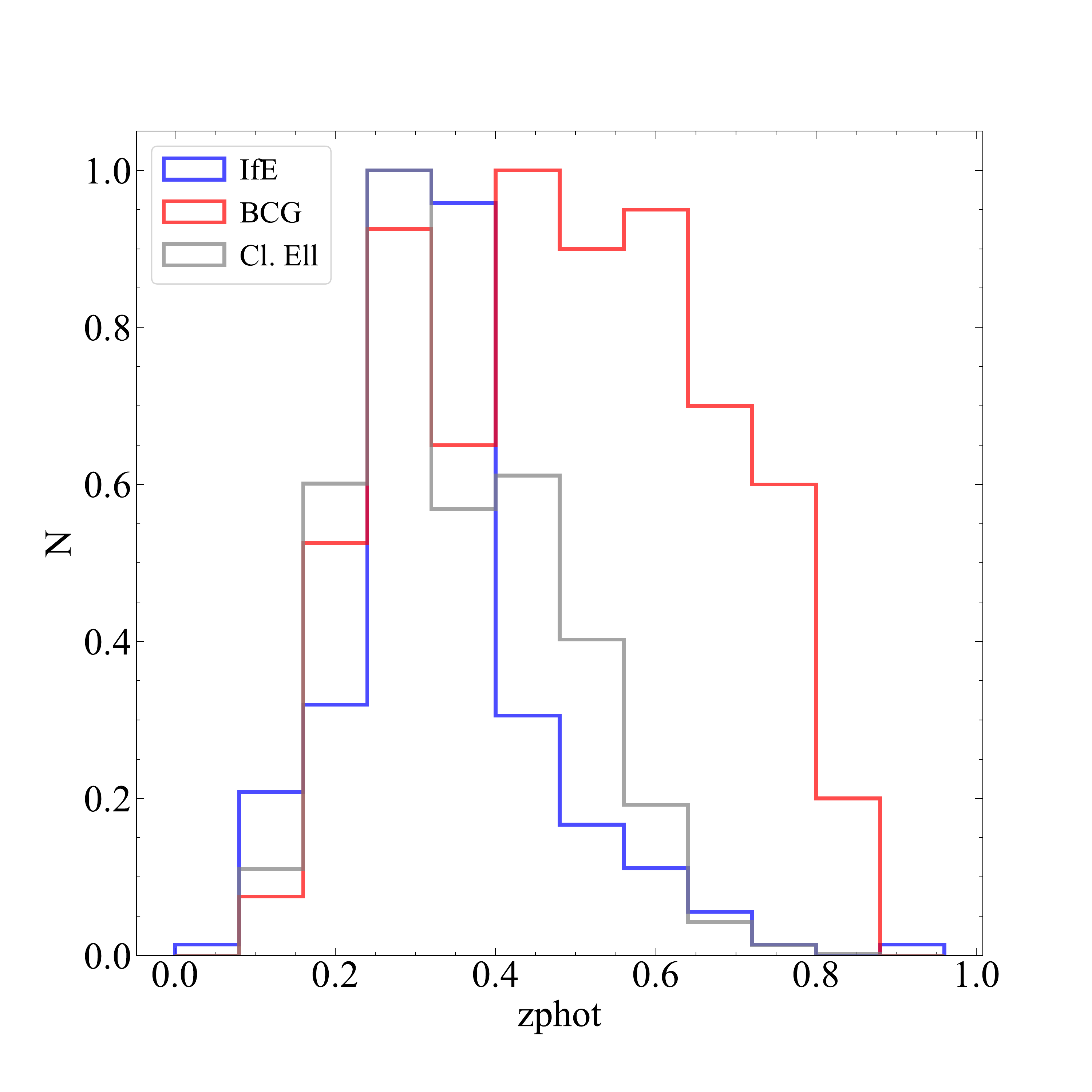}
	\caption{Photometric redshift distribution of our sample galaxies. Sample sizes and redshift ranges are given in Table \ref{tabsample}.}
	\label{histzpall}
\end{center}
\end{figure}

The distribution of the photometric redshifts of our galaxy samples is given in Fig. \ref{histzpall}. Due to their higher intrinsic brightness, BCGs span a relatively larger redshift range with respect to IfEs. The median redshift difference between the BCGs and the ellipticals from the same clusters is comparable with the accuracy of photometric redshifts.

\subsection{Brightness and Colour}

\begin{figure*}[!ht]
\begin{center}
	\includegraphics[width=\columnwidth]{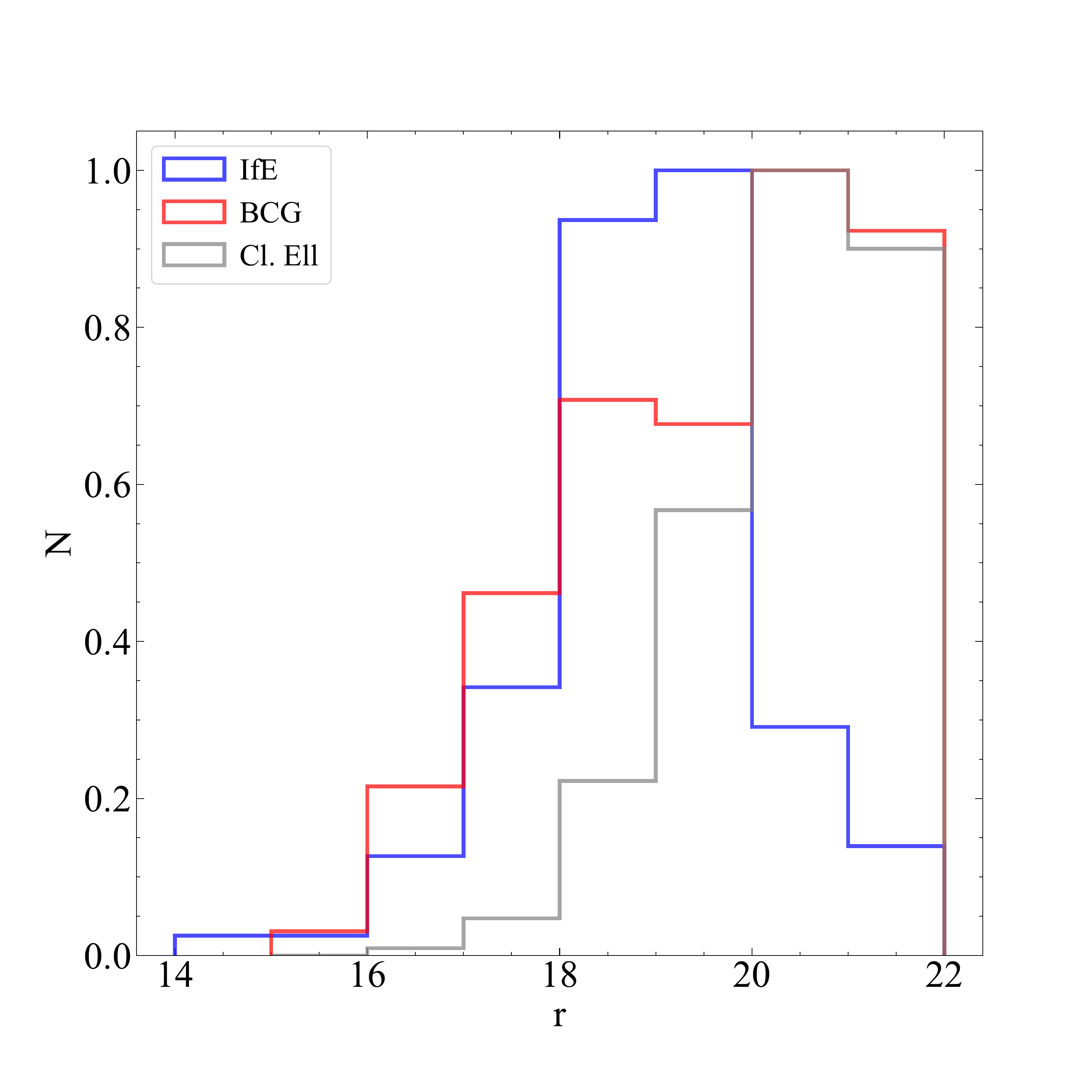} 
	\includegraphics[width=\columnwidth]{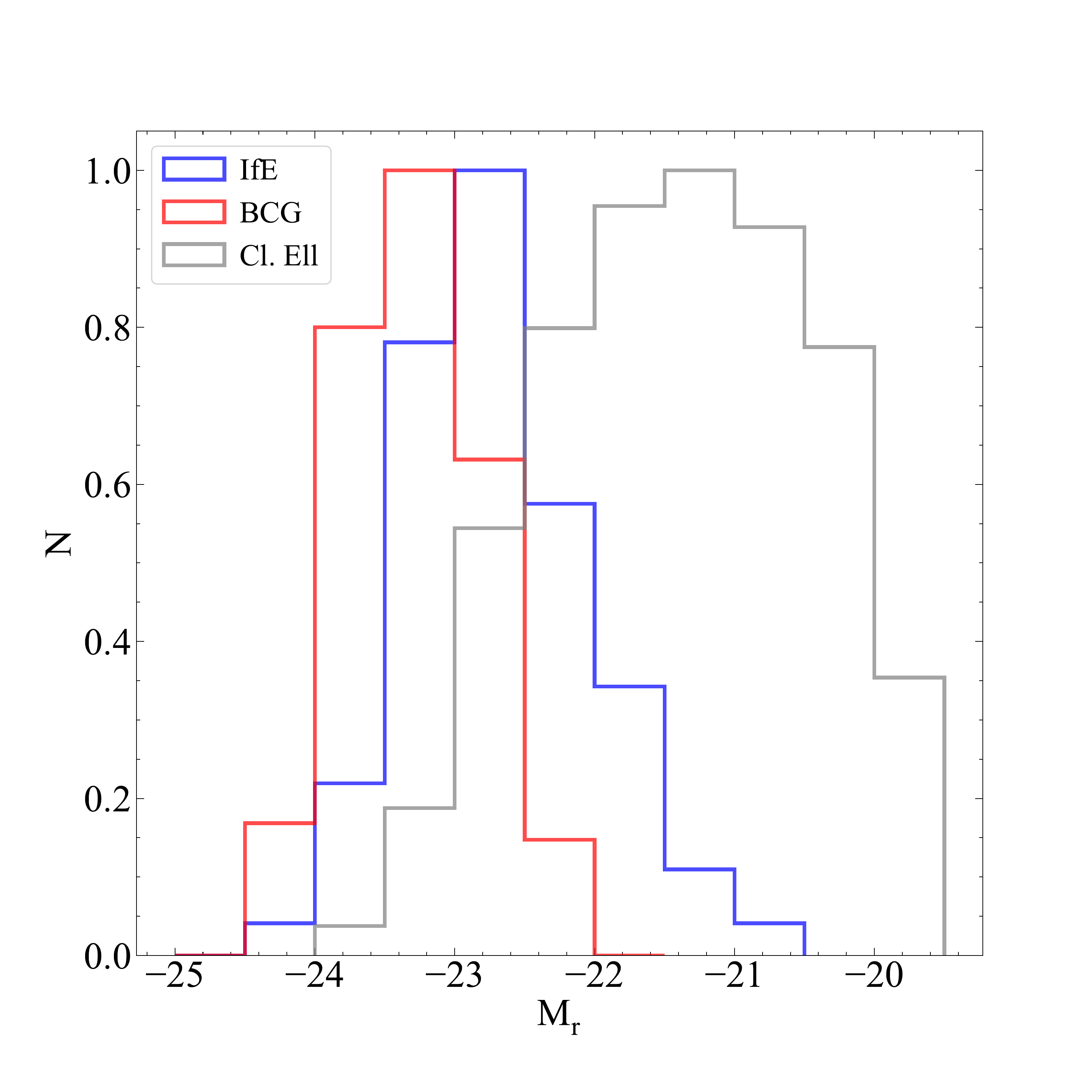}
	\caption{r-band apparent (left) and absolute magnitude (right) distributions of IfEs, BCGs, and cluster ellipticals. Histograms are given as normalized.}
	\label{maggal}
\end{center}
\end{figure*}

Apparent and absolute magnitude distributions of IfEs, BCGs, and cluster ellipticals are given in Fig. \ref{maggal}. Absolute magnitude distribution indicates that IfEs are intrinsically brighter than cluster ellipticals but slightly fainter than BCGs. 

The similarity in absolute magnitudes of IfEs with BCGs may be explained by the scenario of fossil group collapse  \citep{Ponman1994,Jones2000,Reda2004} for their formation as the merging of several galaxies could contribute to the final luminosity of the isolated galaxy.

As our galaxy samples span a large redshift range of 0.1 $\leq$ z $\leq$ 0.9, to for account evolutionary effects, rest-frame colours obtained by \texttt{LePhare} are used for the comparison.

\begin{figure}[!h]
\begin{center}
	\includegraphics[width=\columnwidth]{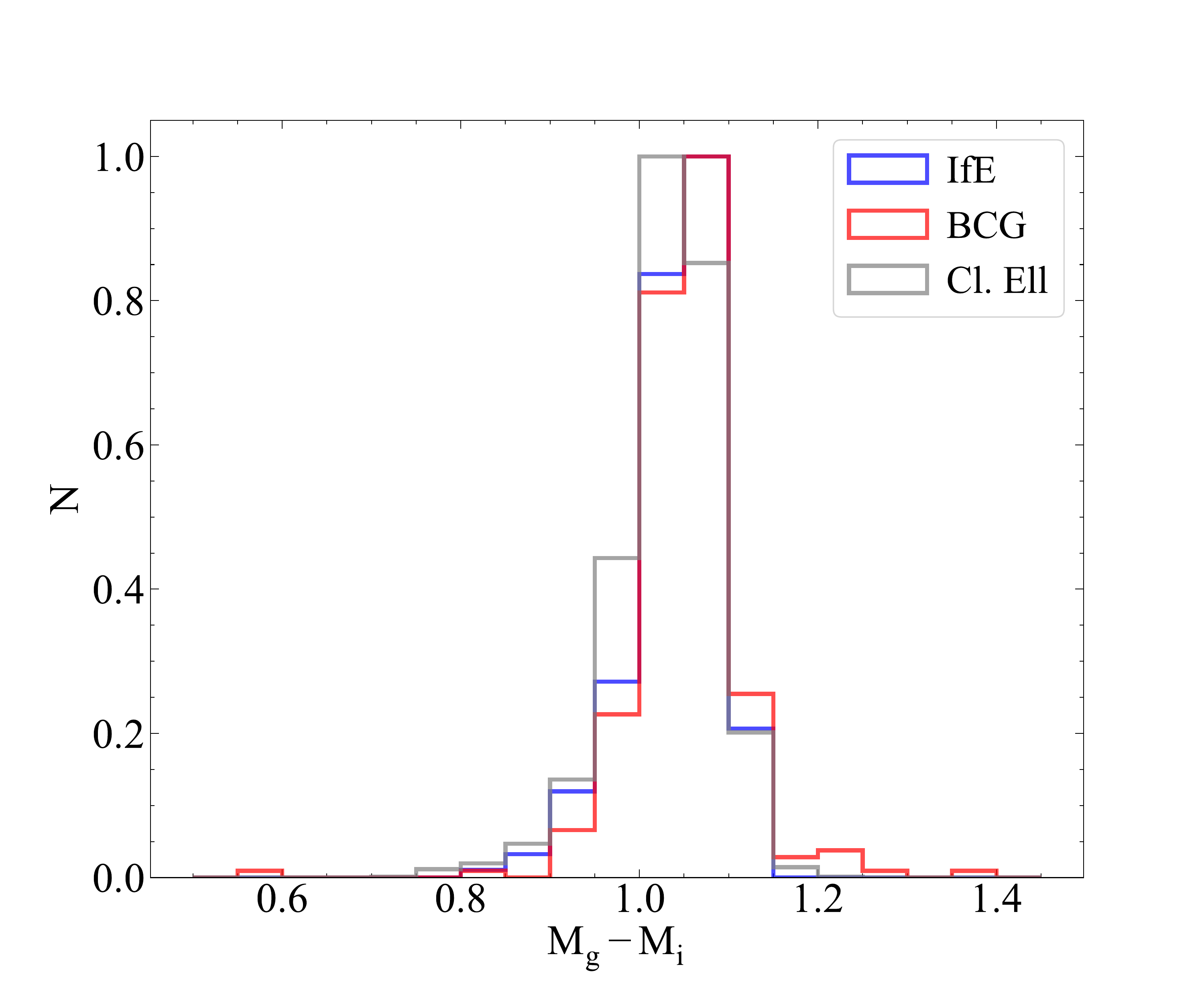}
	\caption{Colour distribution of IfEs, BCGs, and cluster ellipticals. Gray dashed histogram represents BCGs, red dot-dashed histogram represents cluster ellipticals and blue solid histogram represents IfEs.}
	\label{histColor}
\end{center}
\end{figure}

$M_g-M_i$ distributions that are given in Fig. \ref{histColor} look similar for the three groups of galaxies. This can be attributed to their early-type nature where they are expected to concentrate on the so-called red-sequence of the bimodal galaxy colour distribution. However, a two-sample Kolmogorov-Smirnov (K-S) test reveals differences that are statistically significant. We applied the K-S test to our samples with a pairwise approach as (IfEs - BCGs), (IfEs - cluster ellipticals), and (BCGs - cluster ellipticals). p-values of $1.91 \times 10^{-2}$ and $1.27 \times 10^{-7}$ obtained for the IfEs-cluster ellipticals and BCGs-cluster ellipticals, respectively, indicate the colour distribution of both IfEs and BCGs significantly different than cluster ellipticals. However, for the IfE-BCG comparison, a p-value of 0.12 is obtained which suggests we can not differentiate these two groups statistically.

Mean $M_g-M_i$ colours and corresponding standard deviations of the samples are as follows: IfEs (1.039, 0.051), BCGs (1.053, 0.064), and cluster ellipticals (1.031, 0.057). Based on these values we also applied a one-way analysis of variance (ANOVA) test to see whether these three groups have differences in their rest-frame colour distributions. A one-way ANOVA test reveals a p-value of $3.8 \times 10^{-9}$ which also suggests there is a significant difference amongst these samples.

\subsection{Environmental Density}

Due to the isolation criteria used in this study, the environment of IfEs is not totally empty which is also the case in \cite{Niemi2010}. Neighbouring galaxies are the ones that do not affect the IfE selection as described in Section 2.2. Those galaxies are mostly the faint ones as implied by the selection criteria.

We computed the environmental density within the 1 Mpc radius for each IfE and BCG in our sample. Since BCGs and respective cluster ellipticals are in the same environment we did not compute the densities for the cluster ellipticals.

For the density computation, we count galaxies up to \emph{0.4L*} where \emph{L*} is the characteristic luminosity of the galaxy luminosity function. The reason for this upper luminosity limit is to prevent any magnitude bias when we compute the densities. When CFHTLS-W1 limiting magnitude is taken into account, \emph{0.4L*} remains brighter for the redshift range of our study. When counting neighbouring galaxies we impose that they are in the same redshift cylinder as we did in the IfE selection and we used \emph{L*} values for the corresponding redshift.

Among our IfEs we have 37 candidates without any neighbour galaxy down to \emph{0.4L*} within the 1 Mpc projected radius. IfE with the densest environment has 21 neighbouring galaxies which correspond to a projected density of 6.7 gal/Mpc$^2$ (ID\#174).

\begin{figure}[!ht]
\begin{center}
	\includegraphics[width=\columnwidth]{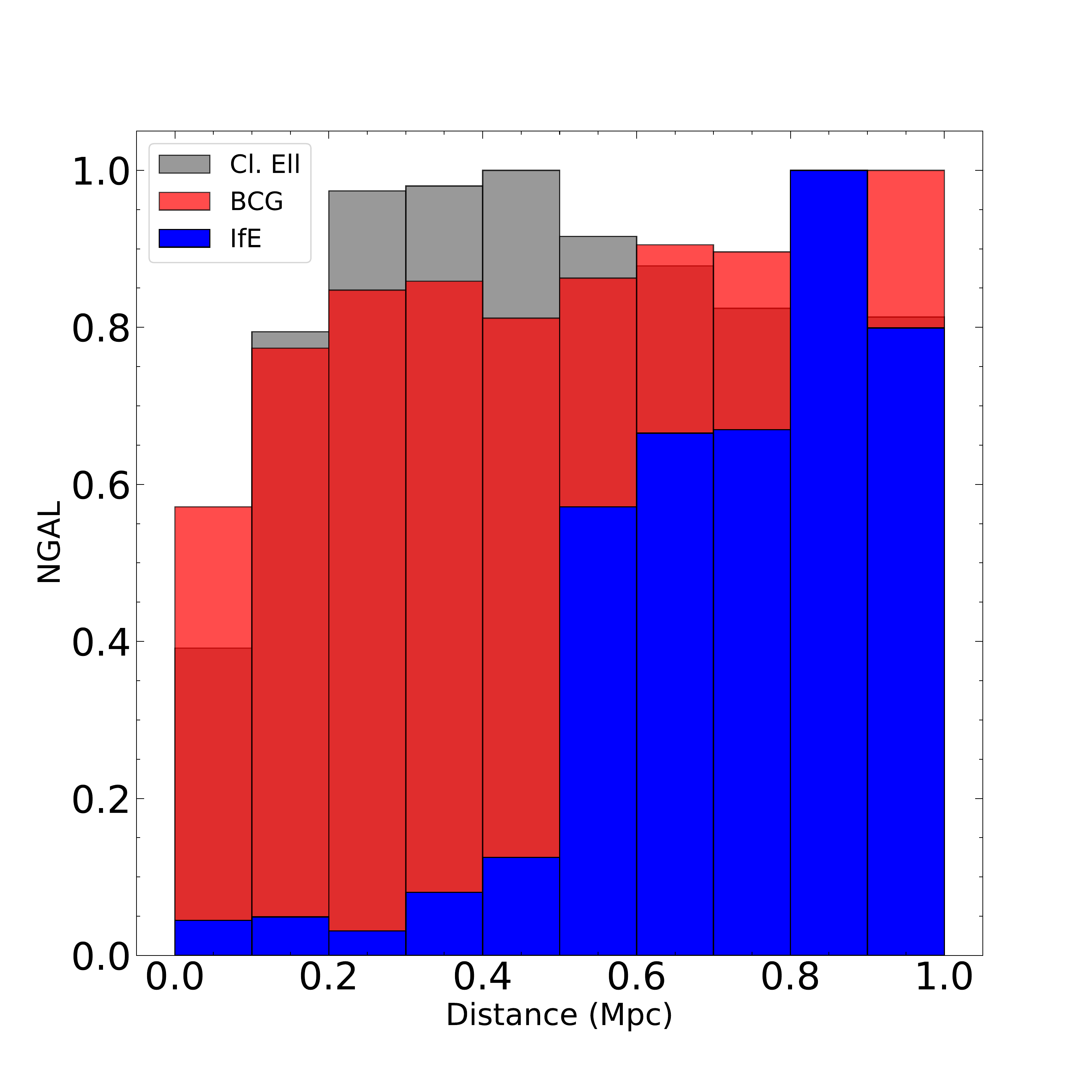}
	\caption{Normalized histogram of neighbouring galaxies (down to \emph{0.4L*}) as a function of distance from parent galaxies. A total of 904 galaxies for IfEs (blue), 10384 galaxies for BCGs (red), and 59779 galaxies for cluster ellipticals (gray) are shown.}
	\label{distance}
\end{center}
\end{figure}

\begin{figure}[!h]
\begin{center}
	\includegraphics[width=\columnwidth]{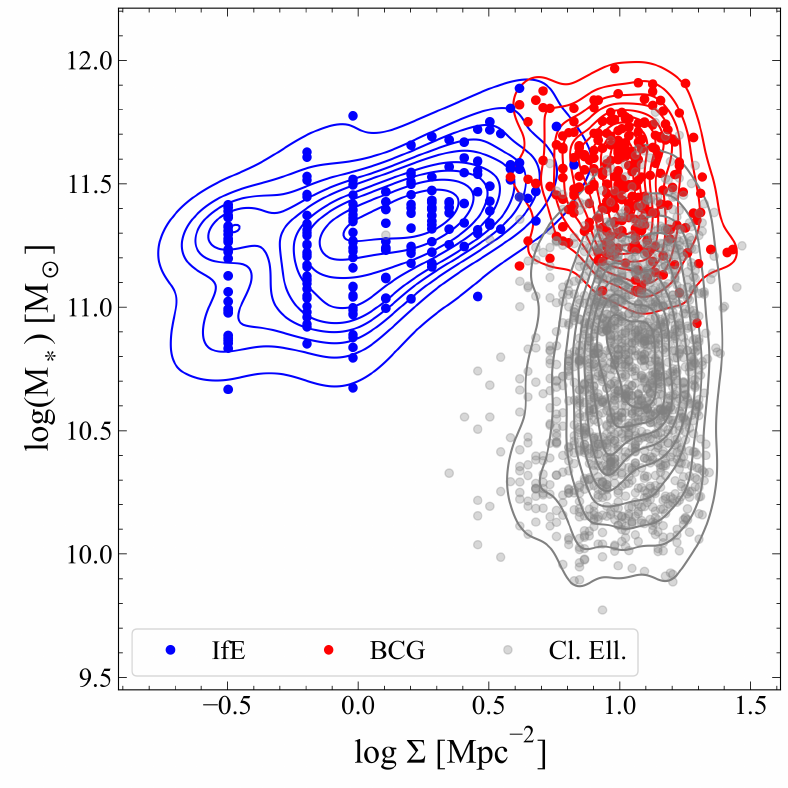}
	\caption{Comparison of stellar masses with corresponding environmental densities for the three samples of galaxies. Densities given here were computed as the projected densities within the 1 Mpc of the galaxy of interest.}
	\label{densitymass}
\end{center}
\end{figure}

The fact that BCGs are located in denser environments can be seen in Fig. \ref{distance}. In the figure, we show the number density of neighbouring galaxies as a function of the distance between our galaxy of interest (e.g. IfE or BCG). Especially for closer distances, such as r$<$0.5 Mpc, densities around BCGs are almost eight times higher than IfEs.

In Fig. \ref{densitymass} we show the stellar mass of IfEs, BCGs, and cluster ellipticals as a function of the local (i.e. 1 Mpc) environmental density. Due to the same number of neighbouring galaxies, some IfEs have discrete density values. However, there seems a trend in the IfE stellar mass with increasing environmental density. This trend does not seem valid for BCGs and cluster ellipticals where the stellar masses of them look almost constant or vary very slowly with density. This may be explained by the cease of the major merging events for BCGs as shown by \cite{DeLucia2007} using the Millennium Simulation. In their study, they showed that 80\% of the stellar masses of BCGs are already formed by z $\sim$ 3. This result can also be considered valid for the other ellipticals in clusters since they are residing in similar environments. The trend for IfEs mainly arises due to the wide range of environments in which they are located. There is approximately a magnitude difference in stellar masses of the IfEs residing in lower and higher densities. This could be the result of merger-driven mass growth for IfEs where such events were identified observationally by \cite{Reda2004} and from simulations by \cite{Niemi2010}. There are no close companion galaxies being merged with IfEs as imposed by the isolation criteria. Thus, morphological disturbances can be investigated for tracing recent or past merging events as we also point out in Fig. \ref{fig2ifes}.

\subsection{Star Formation Activity}

Star formation activity in elliptical galaxies is not very common \citep{Holmberg1958,Conselice14,Madau2014}. However, some elliptical galaxies show excess star formation activities \citep{Pipino2009,Hicks2010,Edman2012,Lacerna2016}. This is mainly related to the environment. For instance in clusters, BCGs may have high star formation rates \citep{Liu2012} due to interaction with galaxies inside the cluster (e.g. cannibalism of dwarf galaxies) \citep{Whiley2008,Lidman2012,Lavoie2016} or the cooling flows towards BCG \citep{Edwards09,McDonald16}.

As $ \textrm H\upalpha$ line is generally the most prominent feature of star-forming galaxies, we compiled $ \textrm H\upalpha$ emission line fluxes to determine whether galaxies in our study show any star formation activity. CFHTLS is an imaging survey, hence we do not have spectra for our galaxies. Thus, we used Sloan Digital Sky Survey (SDSS) Data Release 12 to access the spectral information for our targets. From SDSS DR12, we used the table  $emissionLinesPort$ \citep{Sarzi2006,Cappellari2004,Maraston2011,Thomas2011} to obtain fluxes. Some of our target galaxies are at the limit of a 2.5 m telescope for spectroscopy, therefore we imposed a S/N $\geq$ 3 for reliable results when selecting spectra. In this way, we could obtain $ \textrm H\upalpha$ fluxes for 63 IfEs, 46 BCGs, and 39 cluster ellipticals.

We then used the equation below given by \cite{Kennicutt1998} to determine star formation rates (SFR) using $ \textrm H\upalpha$ line fluxes.

\begin{equation}
	SFR \ (H\alpha) = 7.9 \times 10^{-42} L_{H\alpha} \ M_{\odot} yr^{-1}
\end{equation}
{\vskip 2mm}

$ \textrm H\upalpha$ line fluxes obtained from SDSS were converted to luminosities using redshifts, and their corresponding luminosity distances within the standard $\Lambda$CDM cosmology.

\begin{figure}[!ht]
\begin{center}
	\includegraphics[width=\columnwidth]{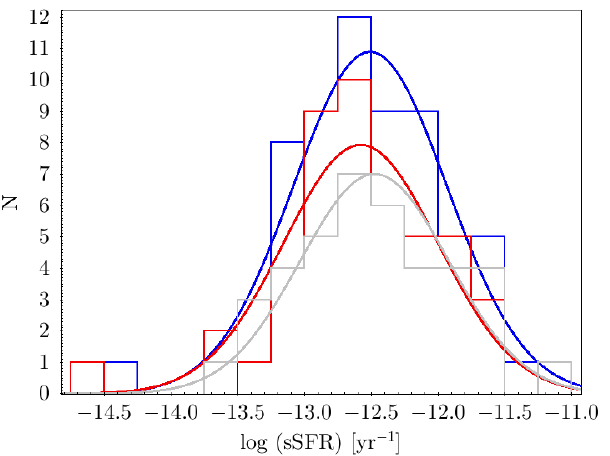}
	\caption{Distribution of specific SFR values for IfEs, BCGs, and cluster ellipticals.}
	\label{ssfr}
\end{center}
\end{figure}

Mean (SFR) for IfEs, BCGs, and cluster ellipticals are as follows: 0.15 $\textrm M_{\odot} \textrm{yr}^{-1}$, 0.18 $\textrm M_{\odot} \textrm{yr}^{-1}$, and 0.14 $ \textrm M_{\odot} \textrm{yr}^{-1}$, respectively.

In order to prevent bias due to large mass differences amongst these samples we also computed the specific star formation rates (sSFR) where sSFR is defined as $ \textrm sSFR \equiv \textrm SFR / \textrm M_{*} $. Fig. \ref{ssfr} shows the distribution of sSFR for IfEs, BCGs, and cluster ellipticals. The distributions were represented with similar Gaussian functions. We also applied K-S and ANOVA tests to see whether there are significant differences between the distributions. However, based on both K-S and ANOVA test results it is difficult to distinguish these samples statistically. The p-value of the ANOVA test was obtained as 0.70 whereas the p-values for K-S tests for the sample pairs are as follows: 0.99 (IfEs-BCGs), 0.97 (IfEs-cluster ellipticals), and 0.93 (BCGs-cluster ellipticals).

\FloatBarrier

\section{Scaling Relations}
\subsection{Size-Luminosity Relation}

In order to check this relation, we make use of effective radii in $r$-band of our sample galaxies as measured by SExtractor and given in the CFHTLS T0007 official catalogue. Using MegaCam pixel scale (i.e. 0.186 "/pixel), they were first converted to angular units (i.e. arcseconds) and then to the physical units (i.e. kpc) by taking into account the photometric redshifts together with the standard cosmological model.

We apply a cut, $-24 \leq M_{r} \leq -22$, to the absolute magnitudes of the three samples to prevent any bias introduced by their distributions. This magnitude cut ensures the largest overlap between the samples. We then divided our samples into four redshift bins as [0.1,0.3], [0.3,0.5], [0.5,0.7], and [0.7,0.9]. However, it is worth noting that there is only one IfE in the last redshift bin. For each redshift bin, we computed the statistics as given in Table \ref{tabsize1}. As it can be expected due to the merger-driven size growth in dense environments \citep{Lidman2013,Lavoie2016}, BCGs are larger than the two other early-type populations in all redshift bins. BCGs are larger than IfEs with a minimum difference of median $ \textrm{log} \textrm R_{ \textrm e}$ is 8\% at $0.7 < \textrm z\leq0.9$ and increases to 23\% at $0.1< \textrm z\leq0.3$. When compared with elliptical galaxies in rich clusters, both BCGs and IfEs are larger. IfEs are larger than cluster ellipticals for the first three redshift bins as 10\%, 4\%, and 8\% for redshift bins with an increasing order (see Table \ref{tabsize1}). Only for the last redshift bin (i.e. $0.7 < \textrm z\leq0.9$) cluster ellipticals 4\% larger than IfEs. We recall the low statistics due to a single IfE in the last bin. Similarly, BCGs are larger than cluster ellipticals as 35\%, 24\%, 20\%, and 5\%.

Afterwards, we applied linear regressions to the size-luminosity relations. Results are obtained for the whole sample without a cut in magnitude but applied separately for each redshift bin. Table \ref{tabsize2} lists coefficients for the individual size-luminosity relations of each galaxy population in each redshift bin as well as the whole redshift range.

\begin{figure}[h]
\begin{center}
	\includegraphics[width=\columnwidth]{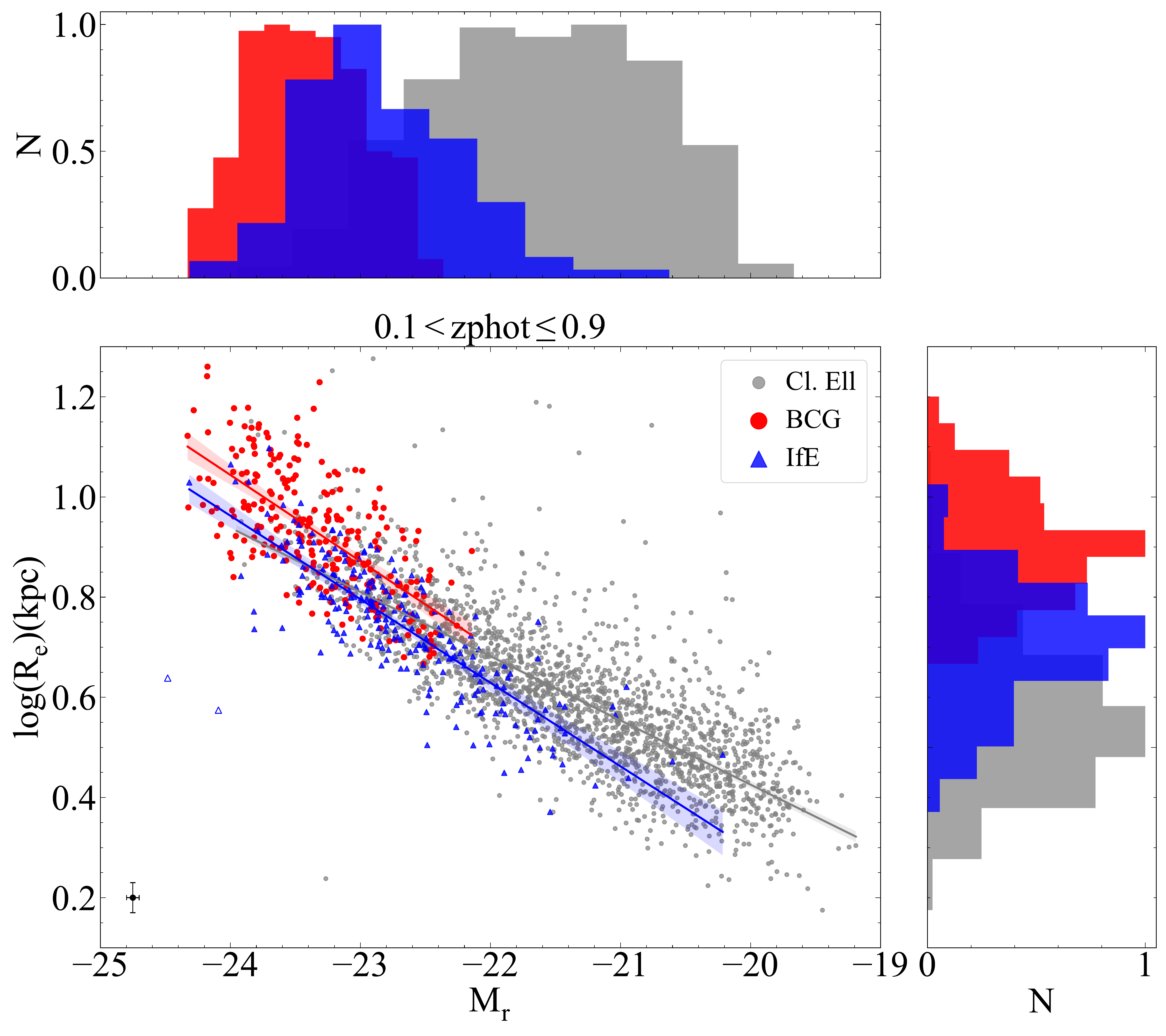} 
	\caption{Absolute magnitude versus effective radius relation. Blue, red and gray points denote IFEs, BCGs and cluster ellipticals, respectively. Similary, blue, red, and gray solid lines are the best fits. Typical errors on both parameters are given as representative at the bottom left of the plot. Statistics of the distributions are given in Table \ref{tabsize1} and coefficients of the linear regressions are given in Table \ref{tabsize2}. For clarity, histograms at the sides are given as normalized.}
	\label{MrlogRe}
\end{center}
\end{figure}

Fig. \ref{MrlogRe} shows the best fits for the three samples. To obtain intrinsic scatter of the three populations, 1000 bootstrap realizations were performed. The resulting confidence intervals are shown with blue, gray, and red shaded areas for IfEs, BCGs, and cluster ellipticals, respectively. The distribution of data and their confidence intervals are also shown in Fig. \ref{MrlogRe}.

\begin{table*}[h!t]
\caption{Statistical properties (size, mean, standart deviation and median) of effective radii (in kpc) and r-band absolute magnitudes for IfEs, BCGs, and cluster ellipticals with a cut in absolute magnitude as $-24 \leq M_{r} \leq -22$ for different redshift bins.}
\centering
\resizebox{\textwidth}{!}{
\begin{tabular}{@{}c|cccc|cccc|cccc|cccc@{}}
\hline
\noalign{\vskip 1mm}
\multicolumn{17}{c}{$logR_{e}$ (kpc)} \\
\noalign{\vskip 1mm}
\hline
\noalign{\vskip 1mm}
\multicolumn{1}{c}{} & \multicolumn{4}{c}{$0.1< \textrm z\leq0.3$} & \multicolumn{4}{c}{$0.3< \textrm z\leq0.5$} & \multicolumn{4}{c}{$0.5< \textrm z\leq0.7$} & \multicolumn{4}{c}{$0.7< \textrm z\leq0.9$}\\
\noalign{\vskip 1mm}
\hline
\noalign{\vskip 1mm}
     & N & Mean & $\sigma$ & Median & N & Mean & $\sigma$ & Median & N & Mean & $\sigma$ & Median & N & Mean & $\sigma$ & Median \\
\noalign{\vskip 1mm}
\hline
 IfE & 56 & 0.74 & 0.12 & 0.75 & 109 & 0.78 & 0.09 & 0.78 & 19 & 0.85 & 0.12 & 0.82 & 1 & 0.84 & - & 0.84 \\
 BCG & 45 & 0.93 & 0.13 & 0.92 & 79 & 0.93 & 0.13 & 0.93 & 94 & 0.90 & 0.11 & 0.91 & 27 & 0.91 & 0.10 & 0.91 \\
 Cl.Ell & 90 & 0.70 & 0.10 & 0.68 & 239 & 0.76 & 0.11 & 0.75 & 247 & 0.78 & 0.10 & 0.76 & 16 & 0.89 & 0.10 & 0.87 \\
\hline
\noalign{\vskip 1mm}
\multicolumn{17}{c}{$M_{r}$ (mag)} \\
\noalign{\vskip 1mm}
\hline
\noalign{\vskip 1mm}
\multicolumn{1}{c}{} & \multicolumn{4}{c}{$0.1< \textrm z\leq0.3$} & \multicolumn{4}{c}{$0.3< \textrm z\leq0.5$} & \multicolumn{4}{c}{$0.5< \textrm z\leq0.7$} & \multicolumn{4}{c}{$0.7< \textrm z\leq0.9$}\\
\noalign{\vskip 1mm}
\hline
\noalign{\vskip 1mm}
     & N & Mean & $\sigma$ & Median & N & Mean & $\sigma$ & Median & N & Mean & $\sigma$ & Median & N & Mean & $\sigma$ & Median\\
\noalign{\vskip 1mm}
\hline
 IfE & 56 & -22.74 & 0.46 & -22.70 & 109 & -22.88 & 0.42 & -22.90 & 19 & -23.17 & 0.45 & -23.20 & 1 & -23.92 & - & -23.92 \\
 BCG & 45 & -23.27 & 0.46 & -23.39 & 79 & -23.15 & 0.44 & -23.17 & 94 & -23.24 & 0.42 & -23.25 & 27 & -23.56 & 0.31 & -23.57 \\
 Cl.Ell & 90 & -22.47 & 0.34 & -22.44 & 239 & -22.48 & 0.38 & -22.39 & 247 & -22.62 & 0.39 & -22.57 & 16 & -23.38 & 0.33 & -23.38 \\
\hline
\end{tabular}
}
\label{tabsize1}
\medskip
\end{table*}

In Fig. \ref{MrlogRe}, IfEs and BCGs have similar size-luminosity relations (i.e. slopes) except for the offsets in their average size and luminosity. However, the slope for the cluster ellipticals is significantly different from those two samples.

Obtained \texttt{$logR_{e} - M_{r}$} relations for IfEs, BCGs, and cluster ellipticals in our sample are given in Eqs. \ref{eqife}, \ref{eqbcg}, and \ref{eqell}, respectively. Equations given below are obtained from the whole redshift range (i.e. $0.1 \leq \textrm z \leq 0.9$).

\begin{equation}
\label{eqife}
\footnotesize
log(R_{e})(IfE) = (-0.167\pm0.009)\times M_{r} + (-3.044\pm0.202)  
\end{equation}

\begin{equation}
\label{eqbcg}
\footnotesize
log(R_{e})(BCG) = (-0.173\pm0.012)\times M_{r} + (-3.106\pm0.274)   
\end{equation}

\begin{equation}
\label{eqell}
\footnotesize
log(R_{e})(Cl.Ell) = (-0.128\pm0.002)\times M_{r} + (-2.138\pm0.048)   
\end{equation}

\begin{table*}[h!]
\caption{Size-luminosity relation best-fitting parameters for different populations in different redshift bins.}
\centering
\begin{tabular}{@{}c|cc|cc|cc|cc|cc@{}}
\hline
\noalign{\vskip 1mm}
\multicolumn{11}{c}{$logR_{e}$ (kpc) = \textbf{\texttt{a}} \ $\times$ \ $M_{r}$ + \textbf{\texttt{b}}} \\
\noalign{\vskip 1mm}
\hline
\noalign{\vskip 1mm}
\multicolumn{1}{c}{} & \multicolumn{2}{c}{All} & \multicolumn{2}{c}{$0.1< \textrm z\leq0.3$} & \multicolumn{2}{c}{$0.3< \textrm z\leq0.5$} & \multicolumn{2}{c}{$0.5< \textrm z\leq0.7$} & \multicolumn{2}{c}{$0.7< \textrm z\leq0.9$}\\
\noalign{\vskip 1mm}
\hline
\noalign{\vskip 1mm}
     & \texttt{a} & \texttt{b} & \texttt{a} & \texttt{b} & \texttt{a} & \texttt{b} & \texttt{a} & \texttt{b} & \texttt{a} & \texttt{b}\\
\noalign{\vskip 1mm}
\hline
 IfE & -0.167 & -3.044 & -0.174 & -3.224 & -0.155 & -2.758 & -0.215 & -4.127 & -0.018 & 0.368 \\
 BCG & -0.173 & -3.106 & -0.221 & -4.221 & -0.199 & -3.676 & -0.165 & -2.942 & -0.181 & -3.361 \\
 Cl.Ell & -0.128 & -2.138 & -0.106 & -1.706 & -0.115 & -1.826 & -0.125 & -2.045 & -0.138 & -2.349 \\
\hline
\end{tabular}
\label{tabsize2}
\medskip
\end{table*}

\subsection{Mass-Size Relation}

Similarly to the size-luminosity relation, we investigate the mass-size relation of each galaxy population separately. Samples are divided into four redshift bins as in the previous section. Masses of our sample galaxies were computed as described in Section 2.5.

To ensure covering the same mass range, we select a sub-sample of our galaxies with masses $10.2< \textrm{log}( \textrm M_{*}/ \textrm M_\odot)\leq12.0$ similarly in \cite{Kelkar15}. The lower limit of this adopted mass range is higher than our mass completeness limits for each redshift bin. 

Sample sizes and statistics on the mass ($\textrm{log}(\textrm M_{*}/ \textrm M_\odot)$) and effective radius ($logR_{e}$) are given in Table \ref{tabmass1} for each redshift bin.

BCGs are more massive than both IfEs and cluster ellipticals in almost all redshift bins. In the last bin (i.e. $0.7 \leq \textrm z \leq 0.9$) IfEs seem larger by 0.02 dex but since there are only two IfEs in this bin results are not totally reliable. Thus, it can be stated that BCGs are the largest in our galaxy samples if we omit this very last redshift bin. The smallest mass difference between BCGs and IfEs occurs at $0.5 \leq \textrm z \leq 0.7$ as 0.07 dex whereas the largest difference at $0.1 \leq \textrm z \leq 0.3$ as 0.44 dex. Both BCGs and IfEs are more massive than cluster ellipticals in all redshift bins. The mass difference between BCGs and cluster ellipticals is 0.08 (1.03) dex at the highest (lowest) redshift bin. The same behaviour is also seen for IfEs when compared with cluster ellipticals, with 0.10 dex being the smallest and 0.59 dex being the largest difference.

\begin{figure}[!ht]
\begin{center}
	\includegraphics[width=\columnwidth]{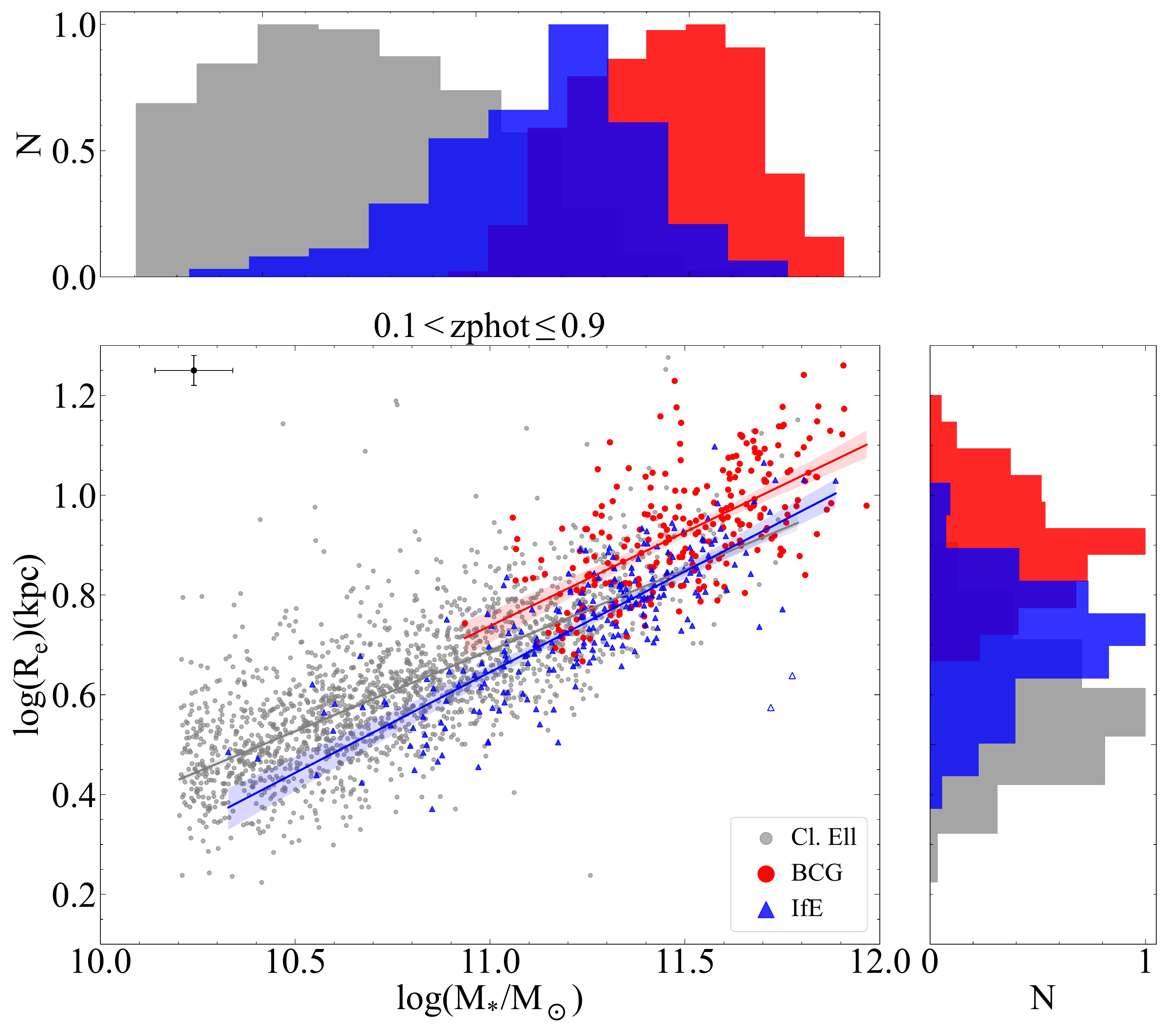} 
	\caption{Stellar mass versus effective radius relation. Blue, red and gray points represent IFEs, BCGs and cluster ellipticals, respectively. Similary, blue, red, and gray solid lines are the best fit for IfEs, BCGs, and cluster ellipticals, respectively. Statistics of the distributions are given in Table \ref{tabmass1} and coefficients of the linear regressions are given in Table \ref{tabmass2}. For clarity, histograms at the sides are given as normalized.}
	\label{masslogRe}
\end{center}
\end{figure}

Mass-size relations for IfEs, BCGs, and cluster ellipticals are shown in Fig. \ref{masslogRe} and corresponding best fits are given in the equations \ref{eqife2}, \ref{eqbcg2}, and \ref{eqell2}. Blue, red, and gray shaded areas in the figure represent the 95\% confidence intervals obtained by bootstrapping 1000 realizations. For the clarity of the figure, errors are not overplotted but a representative error (median) is given in the upper left. BCGs and IfEs have comparable slopes in mass-size relations and are steeper than cluster ellipticals, similarly to the size-luminosity relation given in Fig. \ref{MrlogRe}.

Equations \ref{eqife2}, \ref{eqbcg2}, and \ref{eqell2} obtained for the whole redshift range but coefficients for each redshift bin are given in Table \ref{tabmass2}.

\begin{equation}
\label{eqife2}
\footnotesize
log(R_{e})(IfE) = (0.405\pm0.022)\times {log(M_{*}/M_\odot)} + (-3.802\pm0.247)  
\end{equation}

\begin{equation}
\label{eqbcg2}
\footnotesize
log(R_{e})(BCG) = (0.379\pm0.029)\times {log(M_{*}/M_\odot)} + (-3.437\pm0.331)   
\end{equation}

\begin{equation}
\label{eqell2}
\footnotesize
log(R_{e})(Cl.Ell) = (0.324\pm0.007)\times {log(M_{*}/M_\odot)} + (-2.874\pm0.072)   
\end{equation}

\begin{table*}[h]
\caption{Statistical properties (size, mean, standard deviation and median) of effective radii (in kpc) and mass for IfEs, BCGs, and cluster ellipticals with a cut in stellar masses as $10.2< \textrm{log}(\textrm M_{*}/ \textrm M_\odot)\leq12.0$ from \cite{Kelkar15} for different redshift bins.}
\centering
\resizebox{\textwidth}{!}{
\begin{tabular}{@{}c|cccc|cccc|cccc|cccc@{}}
\hline
\noalign{\vskip 1mm}
\multicolumn{17}{c}{$logR_{e}$ (kpc)} \\
\noalign{\vskip 1mm}
\hline
\noalign{\vskip 1mm}
\multicolumn{1}{c}{} & \multicolumn{4}{c}{$0.1< \textrm z\leq0.3$} & \multicolumn{4}{c}{$0.3< \textrm z\leq0.5$} & \multicolumn{4}{c}{$0.5< \textrm z\leq0.7$} & \multicolumn{4}{c}{$0.7< \textrm z\leq0.9$} \\
\noalign{\vskip 1mm}
\hline
\noalign{\vskip 1mm}
     & N & Mean & $\sigma$ & Median & N & Mean & $\sigma$ & Median & N & Mean & $\sigma$ & Median & N & Mean & $\sigma$ & Median\\
\noalign{\vskip 1mm}
\hline
 IfE & 79 & 0.69 & 0.14 & 0.70 & 124 & 0.76 & 0.12 & 0.77 & 19 & 0.85 & 0.12 & 0.82 & 2 & 0.79 & 0.05 & 0.79 \\
 BCG & 48 & 0.94 & 0.13 & 0.94 & 83 & 0.94 & 0.14 & 0.93 & 97 & 0.90 & 0.11 & 0.91 & 33 & 0.92 & 0.10 & 0.92 \\
 Cl.Ell & 678 & 0.53 & 0.12 & 0.51 & 898 & 0.64 & 0.12 & 0.62 & 306 & 0.76 & 0.10 & 0.74 & 13 & 0.90 & 0.11 & 0.86 \\
\hline
\noalign{\vskip 1mm}
\multicolumn{17}{c}{${log(M_{*}/M_\odot)}$} \\
\noalign{\vskip 1mm}
\hline
\noalign{\vskip 1mm}
\multicolumn{1}{c}{} & \multicolumn{4}{c}{$0.1< \textrm z\leq0.3$} & \multicolumn{4}{c}{$0.3< \textrm z\leq0.5$} & \multicolumn{4}{c}{$0.5< \textrm z\leq0.7$} & \multicolumn{4}{c}{$0.7< \textrm z\leq0.9$}\\
\noalign{\vskip 1mm}
\hline
\noalign{\vskip 1mm}
     & N & Mean & $\sigma$ & Median & N & Mean & $\sigma$ & Median & N & Mean & $\sigma$ & Median & N & Mean & $\sigma$ & Median\\
\noalign{\vskip 1mm}
\hline
 IfE & 79 & 11.17 & 0.28 & 11.17 & 124 & 11.25 & 0.26 & 11.30 & 19 & 11.37 & 0.18 & 11.39 & 2 & 11.68 & 0.01 & 11.68 \\
 BCG & 48 & 11.56 & 0.20 & 11.61 & 83 & 11.46 & 0.21 & 11.45  & 97 & 11.46 & 0.18 & 11.46 & 33 & 11.60 & 0.20 & 11.66 \\
 Cl.Ell & 678 & 10.64 & 0.31 & 10.58 & 898 & 10.77 & 0.30 & 10.75 & 306 & 11.11 & 0.22 & 11.11 & 13 & 11.52 & 0.18 & 11.58 \\
\hline
\end{tabular}
}
\label{tabmass1}
\medskip
\end{table*}

\begin{table*}[h!]
\caption{Mass-size relation best-fitting parameters for different populations in different redshift bins.}
\centering
\begin{tabular}{@{}c|cc|cc|cc|cc|cc@{}}
\hline
\noalign{\vskip 1mm}
\multicolumn{11}{c}{$logR_{e}$ (kpc) = \textbf{\texttt{a}} \ $\times$ \ ${log(M_{*}/M_\odot)}$ + \textbf{\texttt{b}}} \\
\noalign{\vskip 1mm}
\hline
\noalign{\vskip 1mm}
\multicolumn{1}{c}{} & \multicolumn{2}{c}{All} & \multicolumn{2}{c}{$0.1< \textrm z\leq0.3$} & \multicolumn{2}{c}{$0.3< \textrm z\leq0.5$} & \multicolumn{2}{c}{$0.5< \textrm z\leq0.7$} & \multicolumn{2}{c}{$0.7< \textrm z\leq0.9$}\\
\noalign{\vskip 1mm}
\hline
\noalign{\vskip 1mm}
     & \texttt{a} & \texttt{b} & \texttt{a} & \texttt{b} & \texttt{a} & \texttt{b} & \texttt{a} & \texttt{b} & \texttt{a} & \texttt{b}\\
\noalign{\vskip 1mm}
\hline
 IfE & 0.405 & -3.802 & 0.399 & -3.769 & 0.379 & -3.507 & 0.489 & -4.702 & 0.031 & 0.421 \\
 BCG & 0.379 & -3.437 & 0.533 & -5.230 & 0.430 & -3.988 & 0.364 & -3.264 & 0.240 & -1.866 \\
 Cl.Ell & 0.324 & -2.874 & 0.263 & -2.270 & 0.289 & -2.472 & 0.259 & -2.116 & 0.256 & -2.029 \\
\hline
\end{tabular}
\label{tabmass2}
\medskip
\end{table*}

\FloatBarrier

\section{Fossil group connection}

Fossil groups are generally characterized by a dominant central luminous galaxy with an extended bright X-ray gas. These systems contain a few or no $L^{*}$ galaxies other than the central one \citep{Jones2003} and for a given optical luminosity their X-ray luminosities are higher than the corresponding regular groups \citep{Khosroshahi2007}. Regular and symmetric X-ray emission around the central galaxy implies an early formation epoch which also indicates that these systems have not had a merging event recently. The formation of this central galaxy is mostly considered as the first stage of the BCG formation \citep{Jones2003}.

As there are not many galaxies, especially brighter than $L^{*}$, around the central galaxy in fossil groups, isolated elliptical galaxies are considered as links to these systems. The collapse of a galaxy group can produce such isolated bright elliptical galaxies with an extended X-ray emission \citep{Ponman1994}.

The XXL Survey is the deepest and the most extensive X-ray survey carried out with a significant overlap with CFHTLS-W1. This field of the XXL Survey is identified as XXL-North \citep{Pierre2016}. The latest galaxy cluster catalogue, which contains 365 brightest clusters was given by \cite{Adami2018}. This unique dataset obtained in the framework of the XXL Survey, including multi-wavelength follow-up observations, led to identifying fossil group candidates in the CFHTLS-W1 \citep{Adami2018}. Among those candidates, there are three fossil groups inside the W1 field. Those systems are listed in Table \ref{tabfossil} with their XXL cluster ID, coordinates, and spectroscopic redshift.

We investigate these three fossil systems whether they are identified in our algorithm. Matching efforts with various matching radii between 1-10 arcseconds revealed that none of these fossil group candidates have a counterpart in our IfE list. Even though there are various definitions of fossil groups, one can expect to have a central elliptical galaxy with a relatively large magnitude gap with other members \citep{Jones2003}. Therefore, these fossil systems in principle could be identified as isolated field ellipticals. 

\begin{figure}[ht]
\begin{center}
	\includegraphics[width=\columnwidth]{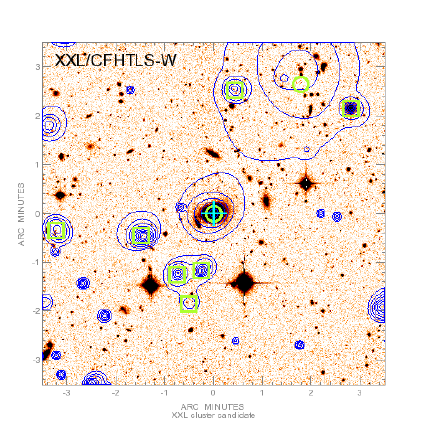}
	\caption{Fossil group candidate XLSSC 147 given by \cite{Adami2018}. X-ray contours are overlaid onto CFHTLS i-band image (Courtesy of XXL Consortium).}
	\label{XLSSC147}
\end{center}
\end{figure}

\begin{figure}[ht]
\begin{center}
	\includegraphics[width=\columnwidth]{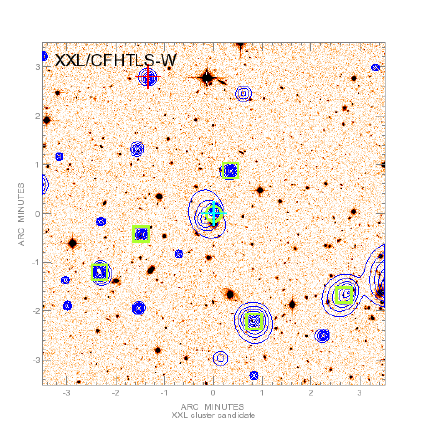}
	\caption{Fossil group candidate XLSSC 162 given by \cite{Adami2018}. X-ray contours are overlaid onto CFHTLS i-band image (Courtesy of XXL Consortium).}
	\label{XLSSC162}
\end{center}
\end{figure}

\begin{figure}[ht]
\begin{center}
	\includegraphics[width=\columnwidth]{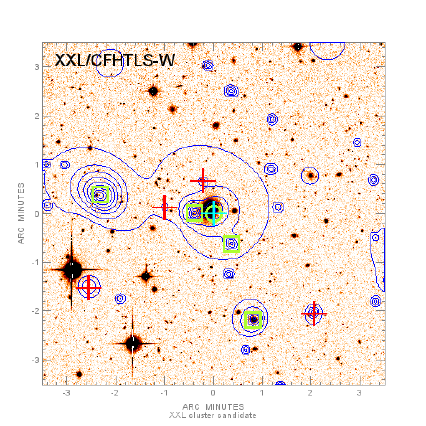}
	\caption{Fossil group candidate XLSSC 171 given by \cite{Adami2018}. X-ray contours are overlaid onto CFHTLS i-band image (Courtesy of XXL Consortium).}
	\label{XLSSC171}
\end{center}
\end{figure}

These fossil group candidates can be seen in Figs. \ref{XLSSC147}, \ref{XLSSC162}, and \ref{XLSSC171} with X-ray contours overlaid. Besides weak diffuse X-ray emission, which is the main reason that \cite{Adami2018} identified these sources as fossil groups, a central bright early-type galaxy is prominent (marked as black triangles in Fig. \ref{positions}). We have checked whether these central elliptical galaxies exist in our input elliptical galaxy catalogue (i.e. 90,872 galaxies) and confirmed their existence. This means that these galaxies do not satisfy the isolation criteria adopted throughout this study. Indeed this is the case and it can be seen from Table \ref{tabfossil2} where we list the neighbour galaxies disturbing the isolation criteria for these central ellipticals.

\begin{table*}[!ht]
\caption{Fossil groups in the CFHTLS-W1 identified using the datasets from the XXL-North \citep{Adami2018} and corresponding central elliptical galaxies. XXL cluster IDs, equatorial coordinates and spectroscopic redshifts for fossil groups; equatorial coordinates, photometric redshift and r-band magnitudes for central elliptical galaxies are given.}
\centering
\begin{tabular}{@{}cccc|cccc@{}}
\hline
\noalign{\vskip 1mm}
\multicolumn{4}{c}{Fossil Group} & \multicolumn{4}{c}{Central Elliptical Galaxy} \\
\noalign{\vskip 1mm}
\hline
\noalign{\vskip 1mm}
XLSSC &   RA      & Dec      &  zspec &  RA & Dec & zphot & r \\
      & (J2000)   & (J2000)  &        &  (J2000) & (J2000) &  &  (mag)\\
\noalign{\vskip 1mm}
\hline
 147 & 37.6409 & -4.6250 & 0.0307 & 37.6410 & -4.6247 & 0.0575 & 14.307 \\
 162 & 32.5239 & -6.0929 & 0.1377 & 32.5234 & -6.0961 & 0.1396 & 17.496 \\
 171 & 31.9860 & -5.8709 & 0.0436 & 31.9868 & -5.8699 & 0.0784 & 15.185 \\
\hline
\end{tabular}
\label{tabfossil}
\medskip
\end{table*}

We also compared our IfE list with all X-ray point sources detected in XXL-North. The 3XLSS point source catalogue is the most comprehensive X-ray catalogue overlapping with CFHTLS-W1 which is published by \cite{Chiappetti2018} and accessible via VizieR online service. The 3XLSS point source catalogue is a product of the \texttt{XAMIN} pipeline of the XXL Survey \citep{Faccioli2018}. The positional accuracy of point sources given by \texttt{XAMIN} pipeline was described in Table 4 of \cite{Chiappetti2013}. According to that table, we adopted a matching radius of 2 arcseconds which is the largest positional error obtained with \texttt{XAMIN}. Only one IfE, ID\#186, has an X-ray counterpart in the 3XLSS catalogue which is marked as a black diamond in Fig. \ref{positions}. This IfE has also been identified as an AGN by SDSS with clear emission lines that can be associated with this behaviour. This IfE has also a spectroscopic redshift of z=0.171 obtained both from SDSS and GAMA Surveys. All other properties of this IfE can be seen in Table \ref{tabife}.

\section{Discussion}

\subsection{Completeness}

Extragalactic studies based on photometric redshifts have the potential to include contamination of foreground/background objects to some degree. In this work, we explore IfEs in the CFTHLS which is a deep imaging survey down to $i\sim 24$ mag for extended objects. 

Searching isolated galaxies based on spectroscopic redshifts would be the ideal approach. The Sloan Digital Sky Survey proved the efficiency of imaging and spectroscopic surveys. However, fibre positioning and moreover survey strategy of the SDSS does not allow searching for isolated galaxies with similar criteria employed in this study, especially beyond the local Universe. \cite{Hernandez2010} and \cite{Argudo-Fernandez15} used SDSS main spectroscopic sample to identify isolated galaxies of all morphological types with $ \textrm z<0.08$. Deeper studies would need dedicated and hence homogeneous spectroscopic surveys such as VIPERS and GAMA.

Due to the limitation of photometric redshifts, completeness in our study is approximately $30\%$ based on the comparison with a spectroscopic sample drawn from GAMA Survey. As it is explained in Section 2.3, we prefer to keep our IfE sample as pure as possible. Thus, we apply a correction on the isolation radius (Case 4 in Table \ref{tabcontrol}) that we employ introduced by the photometric redshift uncertainty. This incompleteness is significant in the conclusions for the highest redshift bin that we explored (i.e. $0.7< \textrm z\leq0.9$).

\subsection{Elliptical Galaxy Colours in Different Environment}

As shown in Section 3.1, the mean colours of IfEs, BCGs, and cluster ellipticals are the same within the 1$\sigma$ standard deviation.

\cite{Lacerna2016} concludes similarly based on their elliptical sample in the local Universe. Based on their comparison of IfEs with Coma supercluster ellipticals, they found that isolated ellipticals and cluster ellipticals have the same colours; (g-i)=1.18 and (g-i)=1.20, respectively. This finding led to the conclusion that most of the ellipticals have red colours and are thus passive (so-called "red and dead") independent from their environment. This result is also shown by \cite{Reda2004} as their isolated galaxies and elliptical galaxies in denser environments have the same average colour. The similarity of the colours of elliptical galaxies implies an early formation epoch where all stars have been born via the transformation of the cold gas.

18 (8\%) IfEs, 16 (6\%) BCGs, and 138 (7\%) cluster ellipticals in our study show bluer colors according to $M_{u}-M_{r}$ separation that we introduced in Section 2.4.1.

Even though a smaller fraction of our samples has bluer colours, we investigated their star formation activity by means of H$\upalpha$ fluxes. However, the sSFR values of sample galaxies are not significantly different from each other. The mean log(sSFR) values for IfEs, BCGs, and cluster ellipticals are as follows -12.51 $\textrm{yr}^{-1}$, -12.58 $\textrm{yr}^{-1}$, and -12.48 $\textrm{yr}^{-1}$, respectively.

\cite{Lacerna2016} also found 8\% of their isolated galaxies as star-forming whereas a significant part of their sample contains bluer IfEs. However, they obtained a mean log(sSFR) for IfEs as -11.88 $\textrm{yr}^{-1}$ and members of Coma cluster as -12.11 $\textrm{yr}^{-1}$. The sSFR value they obtained for IfEs is 0.63 dex larger than our sample.

Star formation activity of the isolated ellipticals might be associated with recent merging events. Such events for the formation scenario of IfEs were proposed by \cite{Reda2004}.

\subsection{Formation Scenarios}

IfEs, BCGs, and cluster ellipticals have all similar colours. This finding implies a common formation mechanism for all elliptical galaxies, irrespective of their environment. At least one could say that the process of morphological transformation of galaxies into ellipticals should be responsible for gas consumption. \cite{Lacerna2016} suggested the SNe Ia feedback via galactic winds or ram-pressure stripping for the removal of gas when the galaxy fly-by of filaments and dense environments. On the other hand, \cite{Reda2004} suggested the major merger of two massive galaxies for the formation of IfEs. In this context, having similar size-luminosity relations shown in Section 4.1 may suggest a similar formation scenario for IfEs and BCGs. The offset between those two relations might be attributed to the difference in corresponding environmental densities.

An alternative scenario has been suggested by \cite{Reda2004} is the collapse of a poor galaxy group. This scenario was proposed due to the high density of dwarf galaxies around IfEs. Such group collapses generally produce strong X-ray sources and/or emissions (i.e. fossil groups). 

An attempt to associate our IfEs with the currently available X-ray group catalogue in CFHTLS-W1 given by \cite{Adami2018} revealed none. Three fossil systems given by \cite{Adami2018} have bright galaxy counterparts in our elliptical galaxy catalogue. If the isolation criteria that we adopt were looser, those fossil systems could have been identified as IfEs. However, in order to be consistent with previous studies, we keep the criteria as it was proposed by \cite{Smith2004}.

As the weak and diffuse X-ray emission from galaxy groups can be disturbed by strong point sources (e.g. AGNs), we have also checked for any counterparts in the X-ray point source catalogue of \cite{Chiappetti2018}. This catalogue contains 14,168 point sources in the 25 deg$^2$ XXL-North field overlapping with CFHTLS-W1. This catalogue of point X-ray sources has been produced with the deepest XMM-Newton observations obtained so far within the framework of the XXL Survey \citep{Pierre2016}. 

The lack or the limited number of X-ray counterparts is consistent with the conclusion of \cite{Niemi2010} where they mentioned some similarities between IfEs and fossil groups but concluded them as two distinct classes. Future X-ray catalogues and IfE samples can help to clarify this formation scenario. Such X-ray catalogues can be provided by the future releases of the XXL Survey or more significantly from the eROSITA All-Sky Survey (i.e. eRASS) which is currently being carried out.

\begin{figure*}[ht!]
\begin{center}
	\includegraphics[width=\columnwidth]{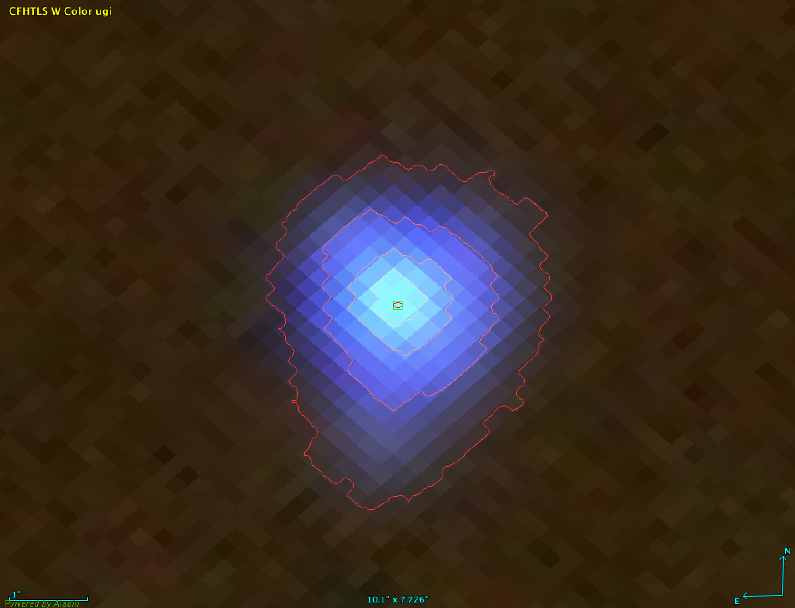}
	\includegraphics[width=\columnwidth]{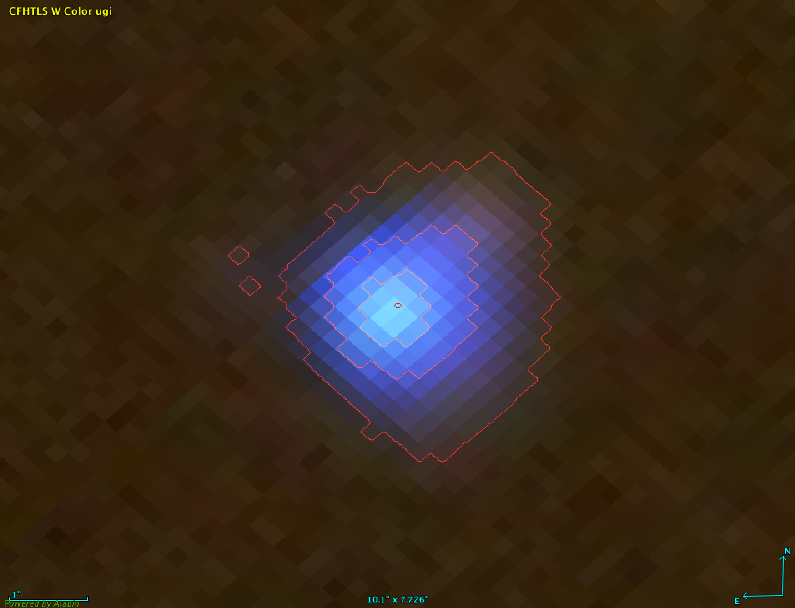}
	\caption{IfEs ID\#119 (z=0.611) (\emph{left}) and ID\#124 (z=0.608) (\emph{right}) showing clear extended emission towards south and north-west directions, respectively. \emph{ugi} three-colour images shown here are from the CFHTLS. Figures and corresponding contours were produced using Aladin applet developed by CDS.}
	\label{fig2ifes}
\end{center}
\end{figure*}

The ideas of group collapse or merger history can be explored via morphological disturbances of galaxies. There are two such IfE candidates in our sample, ID\#119 (z=0.611) and ID\#124 (z=0.608), shown in Fig. \ref{fig2ifes} where their $ugi$ color images from the CFHTLS are given with flux contours overlaid. Light distributions deduced from the contours suggest a disturbed morphology which might indicate a recent merging event. These two galaxies are shown as blue open triangles in Figs. \ref{MrlogRe} and \ref{masslogRe}, took our attention as they are located at the bright end of the IfE sample but they have relatively smaller effective radii than other IfEs.

Such IfEs with disturbed morphologies could be explored via non-parametric structural methods like CAS scheme \citep{Conselice14} or Gini/M20 indices \citep{Lotz04}. Investigating these disturbed morphologies is beyond the scope of this paper and we plan to address this question in an accompanying paper.

\subsection{Scaling Relations}

Comparing elliptical galaxies in different environments may give clues to understanding their formation and evolution. We compare IfEs with BCGs and cluster ellipticals identified from the same survey (i.e. CFHTLS). On average, BCGs are larger and more massive than IfEs and cluster ellipticals. This result is not surprising as BCGs are the most massive and the most luminous galaxies in the Universe \citep{vonderLinden07}. 

Therefore we concentrate on comparing the size-luminosity relation for each galaxy population. In Fig. \ref{MrlogRe}, IfEs and BCGs show similar behaviour with BCGs having an offset towards larger sizes. \cite{Delaye14} also found that the mass-normalized size distribution of cluster early types is skewed towards larger sizes compared with that of the field. They found that the average size of cluster early types is $30-40\%$ larger, while the median size is similar in clusters and in the field. The difference in mean and median sizes given by \cite{Delaye14} is not seen in our sample. For all elliptical galaxies as a whole, the most luminous ones do have larger sizes independent of being BCGs \citep{Bernardi07}.

The slopes of the size-luminosity relations for BCGs and IfEs are similar in our study (see Table \ref{tabsize2}). \cite{Bernardi07} studied 215 BCGs from C4 cluster catalog \citep{Miller05} out to z=0.12. They found that early-type BCGs have a steeper size-luminosity relation than normal early-type galaxies which is in agreement with our results.

\cite{Samir20} compared 2002 BCGs from redMaPPer cluster catalog with  redshifts $0.076< \textrm z<0.394$ and 550 isolated ellipticals from the catalog of SDSS-based Isolated Galaxies (SIG) \citep{Argudo-Fernandez15} with a redshift range of $0.01< \textrm z<0.08$. They found that BCGs follow a steeper size-luminosity relation than isolated ellipticals with a clear tendency where BCGs have larger sizes.

Sizes of IfEs and BCGs can be compared for a certain mass range in order to eliminate any bias that might be introduced due to selection.

\cite{Huertas-Company13B} compared central and satellite galaxies in the mass bins $11< \textrm{log}( \textrm M_{*}/ \textrm M_\odot)\leq11.5$ and $11.5< \textrm{log}( \textrm M_{*}/ \textrm M_\odot)\leq12.0$. They found no difference in mass-size relations for cluster and field population in the local universe (i.e. $ \textrm z<0.09$). According to them, mass-size relation does not depend on the location of the galaxy in the cluster. However, our finding is contradicting as our BCGs have steeper mass-size relation than elliptical galaxies in the same clusters.

Similarly, \cite{Kelkar15} also showed that field and cluster populations have same mass-size relations within the mass range of $10.2< \textrm{log}( \textrm M_{*}/ \textrm M_\odot)\leq12.0$. Our comparison in the same mass range in Section 4.2 shows clear difference between IfEs, BCGs and cluster ellipticals.

\section{Summary and Conclusions}

In this paper, we examined an IfE sample driven from the CFHTLS-W1 galaxy catalogue in the redshift range of $0.07 \leq z \leq 0.91$. We studied the properties of IfEs and compared them with BCGs and cluster ellipticals drawn from the same survey as representing denser environments. The comparison of these samples is performed within similar redshift ranges ($\sim 0.1 \leq z \leq \ \sim 0.9$) and when the same flux limit ($\textrm r < 21.8$) is applied.

We identified 228 IfEs in the 72 sq. deg. field. Those 228 IfEs correspond to a density of $\sim$ 3 IfE/sq. deg. Within our magnitude range, one elliptical galaxy in nearly every 400 is an IfE. In other words, the IfE fraction amongst elliptical galaxies is $0.25$\%. If we take into account all galaxies (regardless of their type) in our working catalogue (i.e. 2,812,065), then this fraction drops to $8.1 \times 10^{-3}\%$, which is a comparable fraction to IfEs in simulated galaxies in the Millennium Simulation \citep{Niemi2010}.

Previous works on IfEs are mostly limited to the local Universe or simulated galaxy catalogues. We identified IfEs up to almost $ \textrm z\sim 1$. Although it was not possible to make a comparison with previous works, there is an overlap with the survey area of \cite{Argudo-Fernandez15}. We identified one common IfE ($\texttt{SDSS J020536.18-081443.1}$) with that study with a spectroscopic redshift of $zspec=0.041$ where the same galaxy has a photometric redshift of $zphot=0.072$ in our catalogue.

The main result of the present study is that IfEs have significant similarities with BCGs. The K-S test based on their colours with a p-value of 0.12 could not distinguish these two samples statistically. Moreover, blue galaxy fractions in both samples are also similar. This is also supported by the very similar sSFR values obtained in this study.

The slope of the mass-size relation is 0.41 and 0.38 for IfEs and BCGs, respectively. For the cluster ellipticals, it is slightly less steep as 0.32. A similar result is obtained for the size-luminosity relation that IfEs and BCGs have nearly the same slope of $-0.17$ where the slope for cluster ellipticals is $-0.13$. Similar trends in these scaling relations suggest a common evolutionary path for IfEs and BCGs.

While it is beyond the scope of the present paper, determining disturbed morphologies of IfEs can provide clues about their origin and possible evolutionary connection with BCGs (e.g. merger driven growth). Randomly, we identified two such IfEs in our catalogue. Such IfEs might support the fossil group collapse scenario for their formation. Image analysis of IfEs based on non-parametric methods (e.g. Gini/M20 indices) could be used to identify such candidates.

Confirmation of the fossil group collapse scenario requires the detection of X-rays from IfEs. Actually, three such IfEs are identified from our sample but due to the isolation criteria that we employed in our study, they were excluded from the final list. However, it would be beneficial to carry-out searches for IfE counterparts in X-rays. The eROSITA All-Sky Survey (eRASS) which is being carried out is the most promising X-ray dataset for such a study.

\section{Acknowledgements}

EKU acknowledges The Scientific and Technological Research Council of Turkey (TUBITAK) BIDEB fellowship through the program 2211-c. This work is supported by Istanbul University with project number T-51339. SA acknowledges the funding by Istanbul University with projects BAP-51339 and BEK-46743. BCG and the cluster elliptical samples of the present work were obtained within the framework of the TUBITAK project 117F311 through the ARDEB-1001 Program.

Based on observations obtained with MegaPrime/MegaCam, a joint project of CFHT and CEA/IRFU, at the Canada-France-Hawaii Telescope (CFHT) which is operated by the National Research Council (NRC) of Canada, the Institut National des Science de l'Univers of the Centre National de la Recherche Scientifique (CNRS) of France, and the University of Hawaii. This work is based in part on data products produced at Terapix available at the Canadian Astronomy Data Centre as part of the Canada-France-Hawaii Telescope Legacy Survey, a collaborative project of NRC and CNRS.

SDSS-III
Funding for SDSS-III has been provided by the Alfred P. Sloan Foundation, the Participating Institutions, the National Science Foundation, and the U.S. Department of Energy Office of Science. The SDSS-III website is \url{http://www.sdss3.org/}.

SDSS-III is managed by the Astrophysical Research Consortium for the Participating Institutions of the SDSS-III Collaboration including the University of Arizona, the Brazilian Participation Group, Brookhaven National Laboratory, Carnegie Mellon University, University of Florida, the French Participation Group, the German Participation Group, Harvard University, the Instituto de Astrofisica de Canarias, the Michigan State/Notre Dame/JINA Participation Group, Johns Hopkins University, Lawrence Berkeley National Laboratory, Max Planck Institute for Astrophysics, Max Planck Institute for Extraterrestrial Physics, New Mexico State University, New York University, Ohio State University, Pennsylvania State University, University of Portsmouth, Princeton University, the Spanish Participation Group, University of Tokyo, University of Utah, Vanderbilt University, University of Virginia, University of Washington, and Yale University.

GAMA is a joint European-Australasian project based around a spectroscopic campaign using the Anglo-Australian Telescope. The GAMA input catalogue is based on data taken from the Sloan Digital Sky Survey and the UKIRT Infrared Deep Sky Survey. Complementary imaging of the GAMA regions is being obtained by a number of independent survey programmes including GALEX MIS, VST KiDS, VISTA VIKING, WISE, Herschel-ATLAS, GMRT, and ASKAP providing UV to radio coverage. GAMA is funded by the STFC (UK), the ARC (Australia), the AAO, and the participating institutions. The GAMA website is \url{http://www.gama-survey.org/}.

XXL is an international project based on an XMM Very Large Programme surveying two 25 deg$^2$ extragalactic fields at a depth of $\sim6x10^{-15} erg s^{-1} cm^{-2}$ in [0.5-2.0] keV. The XXL website is \url{http://irfu.cea.fr/xxl}. Multiband information and spectroscopic follow-up of the X-ray sources are obtained through several survey programmes, summarised at \url{http://xxlmultiwave.pbworks.com/}.

This research has made use of the "Aladin sky atlas" developed at CDS, Strasbourg Observatory, France.

\begin{appendix}

\begin{sidewaystable*}
\begin{center}
\caption{List of isolated field elliptical galaxies. ID, equatorial coordinates, photometric redshift, g-r-i band apparent and absolute magnitudes, effective radius, spectroscopic redshift, redshift source, $ \textrm H{\upalpha}$ flux, star formation rate, and stellar mass are given for each IfE (the last 20 rows). \texttt{The full table can be downloaded from VizieR database.}}
\label{tabife}
\footnotesize\setlength{\tabcolsep}{2.0pt}
\begin{tabular}{c@{\hspace{3pt}}*{1}{cclccccccccccccc}}
\hline
\noalign{\vskip 1mm} 
ID & RA & Dec & zphot & g & r & i & $M_{g}$ & $M_{r}$ & $M_{i}$ & $R_{e}$ & zspec & source & $H_{\alpha}$ flux & SFR & log$M^{*}$ \\
   & \multicolumn{2}{c}{(J2000) (deg)} &  & (mag) & (mag) & (mag) & (mag) & (mag) & (mag) & (kpc) & & &($10^{-17}$ erg $cm^{-2}$ $s^{-1}$) & $M_{\odot}yr^{-1}$ &$M_{\odot}$ \\
\noalign{\vskip 1mm} 
\hline
\noalign{\vskip 1mm}
209   &   34.8577576   &   -4.5408654   &   0.507   &   22.216   &   20.665   &   19.722   &   -21.944   &   -22.598   &   -22.967   &   5.852   &   0.526   &   SDSS   &   -   &   -   &	11.18	\\
210   &   34.4727058   &   -4.4751353   &   0.284   &   19.878   &   18.74   &   18.199   &   -21.755   &   -22.428   &   -22.782   &   5.903   &   0.253   &   GAMA   &   -   &   -   &	11.11	\\
211   &   34.7110977   &   -4.4684629   &   0.291   &   22.388   &   21.037   &   20.492   &   -19.532   &   -20.214   &   -20.538   &   3.057   &   -   &   -   &   -   &   -   &	10.33	\\
212   &   34.0458527   &   -4.4411793   &   0.294   &   19.888   &   18.668   &   18.077   &   -21.931   &   -22.653   &   -22.993   &   6.399   &   0.292   &   GAMA   &   -   &   -   &	11.21	\\
213   &   34.472187   &   -4.1265574   &   0.362   &   20.013   &   18.497   &   17.84   &   -22.758   &   -23.472   &   -23.842   &   8.28   &   0.371   &   SDSS   &   8.927   &   0.213   &	11.61	\\
214   &   34.7439384   &   -4.0216975   &   0.106   &   17.193   &   16.434   &   16.008   &   -21.518   &   -22.132   &   -22.495   &   3.194   &   0.070   &   GAMA   &   -   &   -   &	11.00	\\
215   &   34.1782722   &   -3.9712858   &   0.47   &   21.088   &   19.604   &   18.872   &   -22.578   &   -23.271   &   -23.639   &   6.313   &   0.456   &   GAMA   &   -   &   -   &	11.45	\\
216   &   34.1656189   &   -3.8507514   &   0.337   &   21.037   &   19.65   &   19.056   &   -21.364   &   -22.055   &   -22.424   &   4.127   &   0.354   &   GAMA   &   -   &   -   &	10.98	\\
217   &   35.0816269   &   -4.197022   &   0.351   &   19.768   &   18.567   &   18.044   &   -22.549   &   -23.165   &   -23.532   &   6.835   &   0.315   &   GAMA   &   -   &   -   &	11.28	\\
218   &   35.7317848   &   -3.954145   &   0.338   &   20.637   &   19.26   &   18.571   &   -21.77   &   -22.514   &   -22.883   &   6.473   &   0.348   &   GAMA   &   11.355   &   0.239   &	11.21	\\
219   &   36.0769157   &   -4.4122057   &   0.259   &   19.38   &   18.147   &   17.606   &   -22.068   &   -22.775   &   -23.16   &   6.338   &   0.268   &   GAMA   &   1.706   &   0.021   &	11.34	\\
220   &   36.4951248   &   -4.290565   &   0.467   &   21.485   &   19.926   &   19.162   &   -22.257   &   -22.941   &   -23.309   &   6.572   &   0.511   &   SDSS   &   -   &   -   &	11.33	\\
221   &   36.6401443   &   -4.1426911   &   0.373   &   19.986   &   18.577   &   17.962   &   -22.767   &   -23.467   &   -23.835   &   8.168   &   0.343   &   SDSS   &   -   &   -   &	11.54	\\
222   &   36.1036835   &   -4.073422   &   0.31   &   20.924   &   19.62   &   19.033   &   -21.138   &   -21.847   &   -22.216   &   4.436   &   0.300   &   GAMA   &   -   &   -   &	10.90	\\
223   &   37.0042877   &   -4.630928   &   0.779   &   22.85   &   21.285   &   20.05   &   -23.251   &   -23.918   &   -24.287   &   6.956   &   -   &   -   &   -   &   -   &	11.67	\\
224   &   37.2709465   &   -4.3562112   &   0.58   &   21.549   &   20.31   &   19.468   &   -22.646   &   -23.178   &   -23.545   &   5.333   &   0.624   &   SDSS   &   -   &   -   &	11.35	\\
225   &   37.0726814   &   -4.2714119   &   0.216   &   18.086   &   16.964   &   16.44   &   -22.74   &   -23.458   &   -23.862   &   9.727   &   0.210   &   GAMA   &   4.219   &   0.032   &	11.68	\\
226   &   38.6860847   &   -4.6451521   &   0.538   &   21.518   &   20.127   &   19.295   &   -22.582   &   -23.115   &   -23.482   &   6.143   &   0.629   &   SDSS   &   -   &   -   &	11.26	\\
227   &   38.396553   &   -4.0951343   &   0.356   &   19.596   &   18.347   &   17.789   &   -22.824   &   -23.473   &   -23.841   &   6.519   &   0.301   &   GAMA   &   24.297   &   0.382   &	11.44	\\
228   &   38.0668297   &   -4.0260243   &   0.179   &   17.915   &   16.895   &   16.402   &   -22.301   &   -23.008   &   -23.404   &   7.522   &   0.182   &   GAMA   &   -   &   -   &	11.47	\\
\noalign{\vskip 1mm} 
\hline
\end{tabular}
\end{center}
\end{sidewaystable*}

\begin{table}[ht]
\caption{Galaxies disturbing the isolation criteria for the three elliptical galaxies with a possible fossil group connection. Equatorial coordinates, redshift, magnitude difference between central galaxy ($m_{gal}-m_{cen}$), and the projected distance to the central galaxy are given.}
\centering
\begin{tabular}{@{}ccccc@{}}
\hline
RA      & Dec      &  Redshift & $\Delta$ m & Distance \\
(J2000) & (J2000)  &           &  (mag) &  (Mpc) \\
\noalign{\vskip 1mm}
\hline
\noalign{\vskip 1mm}
\multicolumn{5}{c}{XLSSC: 147} \\
\noalign{\vskip 1mm}
\hline
37.4058266 & -4.8994598 & 0.088 & 1.433 & 0.674 \\
37.8905220 & -4.9110050 & 0.072 & 2.18 & 0.708 \\
37.4617615 & -4.3306589 & 0.065 & 1.416 & 0.643 \\
\hline
\noalign{\vskip 1mm}
\multicolumn{5}{c}{XLSSC: 162} \\
\noalign{\vskip 1mm}
\hline
32.5038452 & -6.2140627 & 0.144&  4.508&  0.831 \\
32.5311966 & -6.1958694 & 0.132&  3.5&  0.696 \\
32.6164513 & -6.1812782 & 0.160&  3.874&  0.874 \\
32.4463997 & -6.1711178 & 0.138&  2.488&  0.746 \\
32.4910088 & -6.1673841 & 0.160&  1.157&  0.544 \\
32.4558372 & -6.1619067 & 0.164&  2.554&  0.654 \\
32.4544411 & -6.1618214 & 0.155&  2.693&  0.661 \\
32.5994949 & -6.1351252 & 0.127&  3.32&  0.592 \\
32.5179863 & -6.1205325 & 0.138&  2.248&  0.174 \\
32.6244240 & -6.0938892 & 0.116&  3.387&  0.699 \\
32.4055328 & -6.0868120 & 0.137&  3.556&  0.818 \\
32.6188316 & -6.0691333 & 0.154&  3.297&  0.686 \\
32.5546303 & -6.0121298 & 0.140&  3.704&  0.623 \\
32.5387878 & -6.0163665 & 0.142&  2.511&  0.565 \\
32.6219711 & -6.0049024 & 0.164&  3.705&  0.931 \\
32.5447350 & -6.0055199 & 0.134&  1.52&  0.647 \\
32.5755882 & -5.9919353 & 0.131&  3.454&  0.810 \\
32.5981064 & -5.9791718 & 0.144&  3.298&  0.964 \\
\hline
\noalign{\vskip 1mm}
\multicolumn{5}{c}{XLSSC: 171} \\
\noalign{\vskip 1mm}
\hline
31.9928436 & -6.0923753 & 0.067&  0.564&  0.725 \\
32.0045547 & -5.9978280 & 0.094&  0.376&  0.421 \\
31.8379269 & -5.9663439 & 0.086&  1.135&  0.575 \\
31.8119850 & -5.9124780 & 0.108&  0.952&  0.583 \\
32.1422958 & -5.8204312 & 0.098&  1.903&  0.529 \\
32.1515999 & -5.7796979 & 0.082&  2.19&  0.609 \\
32.1699219 & -5.6875854 & 0.102&  1.331&  0.839 \\
32.1924896 & -5.6412396 & 0.074&  -0.762&  0.999 \\
\hline
\end{tabular}
\label{tabfossil2}
\medskip
\end{table}

\begin{figure*}[ht]
\centering
    \includegraphics[width=17cm]{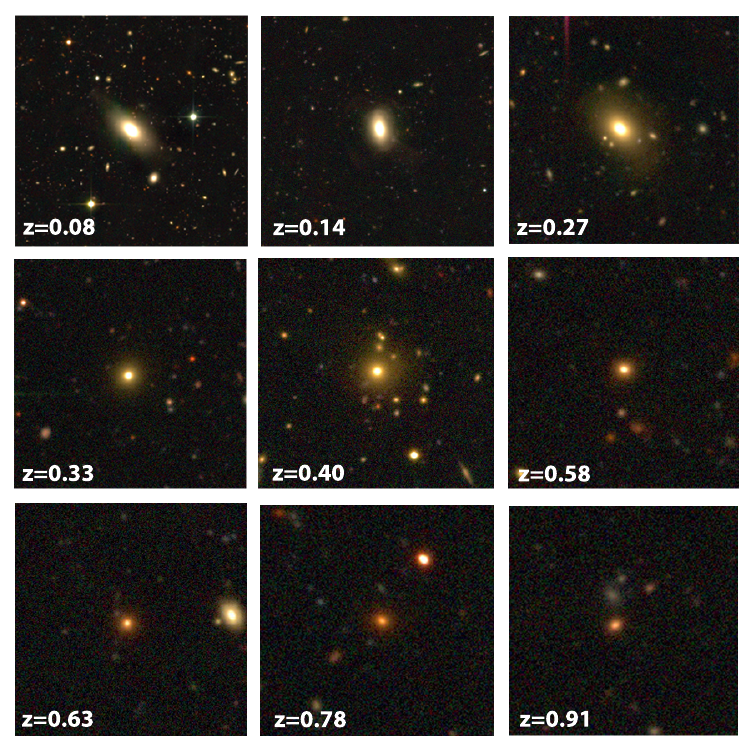}
    	\caption{A sample of IfEs identified in this study. Cutout images are produced from g, r, and i-band images using STIFF. Objects are ordered with increasing redshift. Redshifts are denoted on the cutout images.}%
    \label{Cutout}%
\end{figure*}

\end{appendix}

\bibliographystyle{pasa-mnras}
\bibliography{ulgen.bib}

\end{document}